\documentclass[12pt]{article}
\usepackage[a4paper]{geometry}
\usepackage[italian,english]{babel}
\usepackage[T1]{fontenc}

\usepackage{amsmath, accents}
\usepackage{amsfonts}
\usepackage{amstext}
\usepackage{amssymb}
\usepackage{amsthm}
\usepackage{amscd}
\usepackage{mathrsfs}
\usepackage{bbold}
\usepackage{dsfont}
\usepackage{bbm}

\usepackage[pagebackref,draft=false]{hyperref}
\hypersetup{colorlinks,
linkcolor=myrefcolor,
citecolor=mycitecolor,
urlcolor=myurlcolor}

\usepackage[capitalize]{cleveref}
\usepackage{cite}
\usepackage{caption}
\usepackage{etaremune}

\usepackage{xcolor}
\definecolor{myurlcolor}{rgb}{0,0,0.4}
\definecolor{mycitecolor}{rgb}{0,0.5,0}
\definecolor{myrefcolor}{rgb}{0.5,0,0}
\usepackage{graphicx}
\usepackage{tikz}
\usepackage{tikz-cd}
\usetikzlibrary{graphs,decorations.pathmorphing,decorations.markings}

\usepackage{lipsum}
\usepackage{etoolbox}
\usepackage{makeidx}
\usepackage{sectsty}
\usepackage{dsfont}
\usepackage{enumitem} 
\usepackage[]{latexsym}
\usepackage{braket}
\usepackage{caption}
\usepackage[utf8]{inputenx}
\usepackage{lmodern}
\usepackage{textcomp}
\usepackage{microtype}
\usepackage{totcount}
\usepackage{blindtext}

\newtheorem{theorem}{Theorem}[section]
\newtheorem{remark}[theorem]{Remark}

\newtheorem{proposition}[theorem]{Proposition}

\newtheorem*{proof*}{Proof}

\newcommand{\be}{\begin{equation}}
\newcommand{\ee}{\end{equation}}
\newcommand{\bea}{\begin{eqnarray}}
\newcommand{\eea}{\end{eqnarray}}

\newcommand{\ac}{\mathscr{S}}

\newcommand{\m}{\mathscr{M}}

\newcommand{\pe}{\mathcal{P}(\mathbb{E})}

\newcommand{\fpe}{\mathcal{F}_{\mathcal{P}(\mathbb{E})}}

\newcommand{\fe}{\mathcal{F}_{\mathbb{E}}}

\newcommand{\pssos}{\Pi^\star_\Sigma \Omega^\Sigma}
\newcommand{\os}{\Omega^\Sigma}

\newcommand{\elag}{\mathcal{E}\mathscr{L}}

\newcommand{\dd}{{\rm d}}
\newcommand{\de}{\partial}

\title{The geometry of the solution space of first order Hamiltonian field theories II: non-Abelian  gauge theories}

\author{F. M. Ciaglia$^{2,6}$ \href{https://orcid.org/0000-0002-8987-1181}{\includegraphics[scale=0.7]{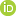}}, F. Di Cosmo$^{1,2,7}$ \href{https://orcid.org/0000-0003-0256-5913}{\includegraphics[scale=0.7]{ORCID.png}}, A. Ibort$^{1,2,8}$ \href{https://orcid.org/0000-0002-0580-5858}{\includegraphics[scale=0.7]{ORCID.png}}, \\ G. Marmo$^{3,4,9}$ \href{https://orcid.org/0000-0003-2662-2193}{\includegraphics[scale=0.7]{ORCID.png}}, L. Schiavone$^{3,5,10}$  \href{https://orcid.org/0000-0002-1817-5752}{\includegraphics[scale=0.7]{ORCID.png}}, A. Zampini$^{3,5,11}$ \href{https://orcid.org/0000-0003-0980-6003}{\includegraphics[scale=0.7]{ORCID.png}} \\
\footnotesize{$^{1}$\textit{ ICMAT, Instituto de Ciencias Matem\'{a}ticas (CSIC-UAM-UC3M-UCM)}} \\
\footnotesize{$^{2}$\textit{Depto. de Matem\'aticas, Univ. Carlos III de Madrid, Legan\'es, Madrid, Spain}} \\
\footnotesize{$^{3}$\textit{ INFN-Sezione di Napoli, Naples, Italy}} \\
\footnotesize{$^{4}$\textit{ Dipartimento di Fisica ``E. Pancini'', Universit\`a di Napoli Federico II,  Naples, Italy}} \\
\footnotesize{$^{5}$\textit{ Dipartimento di Matematica e Applicazioni "Renato Caccioppoli", Università di Napoli Federico II, Napoli, Italy}} \\
\footnotesize{$^{6}$\textit{ e-mail: \texttt{fciaglia[at]math.uc3m.es
}}} \,\, 
\footnotesize{$^{7}$\textit{ e-mail: \texttt{fcosmo[at]math.uc3m.es}}} \\
\footnotesize{$^{8}$\textit{ e-mail: \texttt{albertoi[at]math.uc3m.es}}} \,\,  
\footnotesize{$^{9}$\textit{ e-mail: \texttt{marmo[at]na.infn.it}}} \\ 
\footnotesize{$^{10}$\textit{ e-mail: \texttt{luca.schiavone[at]unina.it}}} \,\,  
\footnotesize{$^{11}$\textit{ e-mail: \texttt{azampini[at]na.infn.it}}} 
}

\begin{document}

\maketitle

\tableofcontents

\begin{abstract}
We go on with the program started in the companion paper \cite{Ciaglia-DC-Ibort-Marmo-Schiav-Zamp2021-Cov_brackets_toappear} of defining a Poisson bracket structure on the space of solutions of the equations of motion of first order Hamiltonian field theories.
The case of non-Abelian gauge theories is addressed by using a suitable version of the coisotropic embedding theorem.
\end{abstract}

\section*{Introduction}
\label{Sec:Intro}
\addcontentsline{toc}{section}{\nameref{Sec:Intro}}

In the first part of this series of papers \cite{Ciaglia-DC-Ibort-Marmo-Schiav-Zamp2021-Cov_brackets_toappear} and in \cite{Ciaglia-DC-Ibort-Marmo-Schiav2020-Jacobi_Particles, Ciaglia-DC-Ibort-Marmo-Schiav2020-Jacobi_Fields} the authors started the analysis of the construction of a covariant Poisson brackets for first order Hamiltonian classical field theories.
The philosophy of the paper was that of equipping the space of solutions of the equations of motion, which is the space capturing the covariance (w.r.t. the relativity group of the theory) properties of the theory, with a Poisson structure.
In particular the authors showed that, within the multisymplectic formulation of first order Hamiltonian field theories, such a space, always referred to as \textit{solution space}, can be canonically equipped with a pre-symplectic structure.
In particular such a structure revealed to be symplectic within those theory non exhibiting any gauge invariance since its kernel is made exactly by the generators of gauge transformations.
For theories without gauge symmetries the Poisson bracket was defined in the usual way by constructing the Poisson tensor as the inverse of the symplectic structure.
Moreover, in \cite{Ciaglia-DC-Ibort-Marmo-Schiav-Zamp2021-Cov_brackets_toappear} the first easiest example of gauge theory was analysed, i.e. Electrodynamics.
In that case the authors, following an idea in \cite{Dubrovin-Giord-Marmo-Sim1993-Poisson_presymplectic}, constructed the Poisson bracket by the aid of a flat connection that could be fixed on a particular bundle associated with the theory.
As already anticipated in \cite{Ciaglia-DC-Ibort-Marmo-Schiav-Zamp2021-Cov_brackets_toappear} such idea does not work for all gauge theories since for some theories, such as Yang-Mills theories, for topological reasons a flat connection of the type used in Electrodynamics can not be fixed.

However, in the present paper we will show that, within the general formulation of first order Hamiltonian field theories presented in \cite{Ciaglia-DC-Ibort-Marmo-Schiav-Zamp2021-Cov_brackets_toappear}, the so called \textit{coisotropic embedding theorem} can be used as a tool to define a Poisson bracket even in cases where a flat connection can not be fixed.
The price one should pay is that the Poisson bracket will be defined on an enlargement of the solution space where additional degrees of freedom appear.
In the case of Yang-Mills theories, which is the main example we will deal with in this paper, the number of additional degrees of freedom emerging via the procedure described turns to be equal to the dimension of $\mathbf{T}^\star \mathfrak{g} \, \simeq \, \mathfrak{g} \times \mathfrak{g}^\star$ ($\mathfrak{g}$ denoting the Lie algebra of the Lie group associated with the gauge theory) and, therefore, can be interpreted as the \textit{ghost} and \textit{antighost} (or \textit{ghost momenta} following the terminology of \cite{Henneaux-Teitelboim1994-Gauge}) appearing in the BRST approach to the quantization of gauge theories.

The paper is organized as follows.
In Sect. \ref{Sec:The coisotropic embedding theorem and Poisson brackets} we recall the content of the \textit{coisotropic embedding theorem}.
In particular we recall its original formulation in Sect. \ref{Subsubsec:Classical coisotropic embedding theorem} and then we give a proof based on the use of a connection in \ref{Subsubsec:The coisotropic embedding theorem via connections}.
Then we argue how to use such a theorem to construct a Poisson bracket on a pre-symplectic manifold, distinguishing the three different cases where the connection used is closed (Sect. \ref{Subsubsec:The closed case}), horizontally closed, i.e. flat (Sect. \ref{Subsubsec:Zero-curvature case}) and non-closed (Sect. \ref{Subsubsec:The non-closed case}).
We will see that in the closed and horizontally closed cases, the Poisson bracket defined on the enlarged space emerging from the coisotropic embedding theorem can be reduced to a Poisson bracket on the original manifold, the solution space.
In particular, it will turn out that the Poisson bivector field on the enlarged space in terms of which the Poisson bracket is defined, can be projected to a Poisson bivector field on the solution space.

Then, since the pre-symplectic manifold we want to analyse is the solution space of Yang-Mills theories, we devote Sect. \ref{Sec:Multisymplectic formulation of Classical Field Theories} to recall, very briefly and avoiding all the proofs, the multisymplectic formulation of first order Hamiltonian field theories extensively discussed in \cite{Ciaglia-DC-Ibort-Marmo-Schiav-Zamp2021-Cov_brackets_toappear} and reference therein.

In Sect. \ref{Sec:Covariant Poisson bracket on the solution space of Yang-Mills theories} the main aim of the paper is addressed, that is, the construction of the Poisson bracket on the solution space of Yang-Mills theories.
In particular, we first address in Sect. \ref{Subsec:A preliminar example: the magnetic monopole} a preliminar (finite-dimensional) example, the magnetic monopole, in order to introduce the construction in a more manageable setting.
Then, in section \ref{Subsec:A second example: free Electrodynamics} we deal with the example of free Electrodynamics already addressed in \cite{Ciaglia-DC-Ibort-Marmo-Schiav-Zamp2021-Cov_brackets_toappear} within the approach in terms of the coisotropic embedding theorem presented in this paper.
In particular, we will show that free Electrodynamics lies in what we called above the "closed case" and, thus, the Poisson bracket defined on the enlarged space obtained via the coisotropic embedding theorem can be projected to the solution space giving rise exactly to the Poisson bracket written in \cite{Ciaglia-DC-Ibort-Marmo-Schiav-Zamp2021-Cov_brackets_toappear}.
Finally, we proceed with the main example of the paper, that is Yang-Mills theories, in Sect. \ref{Subsec:Multisymplectic formulation of Yang-Mills theories} and \ref{Subsec:Poisson bracket on the solution space via coisotropic embedding}.
In particular, in Sect. \ref{Subsec:Multisymplectic formulation of Yang-Mills theories} we recall the multisymplectic formulation of Yang-Mills theories, whereas in Sect. \ref{Subsec:Poisson bracket on the solution space via coisotropic embedding} we proceed with the construction of the Poisson bracket via the coisotropic embedding theorem.


\section{The coisotropic embedding theorem and Poisson brackets}
\label{Sec:The coisotropic embedding theorem and Poisson brackets}

In this section we show how the coisotropic embedding theorem can be seen as a tool to define a Poisson bracket on a suitable enlargement of a pre-symplectic manifold and under which geometrical conditions such a Poisson bracket can be "projected" to a Poisson bracket on the original pre-symplectic manifold.

\subsection{The coisotropic embedding theorem}
\label{Subsec:The coisotropic embedding theorem}

We devote this section to recall the content of the \textit{coisotropic embedding theorem}.
Consider a pre-symplectic manifold $(\mathcal{M},\, \omega)$.
The coisotropic embedding theorem gives a canonical way of embedding $(\mathcal{M},\, \omega)$ into a symplectic manifold, say $(\tilde{\mathcal{M}},\, \Omega)$ such that $(\mathcal{M},\, \omega)$ is a coisotropic submanifold, i. e., $\mathfrak{i}^\star \Omega \,=\, \omega$, $\mathfrak{i}$ denoting the embedding map.

In Sect. \ref{Subsubsec:Classical coisotropic embedding theorem} we sketch the proof of the theorem in its standard formulation, which is due to \textit{M. Gotay} \cite{Gotay1982-Coisotropic_embedding}.
We also refer to \cite{Guillemin-Sternberg1990-Symplectic_techniques} for the proof of the theorem and for its equivariant version.

The original formulation is to some extent too general for our purposes and not well suited for calculations.
For this reason, in Sect. \ref{Subsubsec:The coisotropic embedding theorem via connections} we use a proof of the theorem based on the use of an additional structure, i.e., a connection which makes the theorem less general but better suited to our applications.


\subsubsection{Classical coisotropic embedding theorem}
\label{Subsubsec:Classical coisotropic embedding theorem}

Assume $\omega$ to be of constant rank and denote by $K_m$ the kernel of $\omega$ at $m \in \mathcal{M}$.
Denote by $\mathbf{K}$ the characteristic bundle over $\mathcal{M}$, i. e., the subbundle of $\mathbf{T}\mathcal{M}$ with typical fibre $K_m$.
Its dual bundle, say $\mathbf{K}^\star$, is again a vector bundle over $\mathcal{M}$.
The base manifold $\mathcal{M}$ can be immersed into $\mathbf{K}^\star$ as the zero section, $\sigma_0$, of the fibre bundle.
Along the image of $\sigma_0$ the tangent space to $\mathbf{K}^\star$ canonically splits as $\mathbf{T}_{\sigma_0(m)}\mathbf{K}^\star \,=\, \mathbf{T}_m\mathcal{M} \oplus K^\star_m$.
\begin{remark}
This is true only along the image of $\sigma_0$, since as far as one moves away from it, only the vertical vectors of $\mathbf{K}^\star$ are canonically defined whereas their complement at each point should be specified via an additional choice.
\end{remark}
Consider a complement of $K_m$ into $\mathbf{T}_m\mathcal{M}$, say $W_m$.
Then $\mathbf{T}_{\sigma_0(m)}\mathbf{K}^\star \,=\, W_m \oplus K_m \oplus K^\star_m$.
Therefore, using the fact that $K_m \oplus K^\star_m$ is a symplectic vector space\footnote{The symplectic structure is given by the natural pairing between $K_m$ and its dual, $\omega_{K_m \oplus K_m^\star}(V+ \rho,\, V' + \rho') \,=\, \langle \, \rho \,,\, V'\, \rangle - \langle \, \rho' \,,\, V \, \rangle$ where $V, V' \in K_m$ and $\rho$, $\rho' \in K_m^\star$.} and that the original pre-symplectic structure $\omega$ is non-degenerate when contracted along elements of $W_m$, a symplectic structure can be constructed on $\sigma_0(\mathcal{M})$ in the following way:
\be
\Omega_0 \,=\, \tau^\star \omega + \omega_{\mathbf{K}\oplus \mathbf{K}^\star} \circ \mathrm{pr} \,,
\ee
where $\tau$ is the projection of the fibre bundle $\mathbf{K}^\star \to \mathcal{M}$ and $\mathrm{pr}$ is the projection $\mathbf{T}_{\sigma_0(m)}\mathbf{K}^\star \to K_m \oplus K_m^\star$.
The extension of $\Omega_0$ to the whole $\mathbf{K}^\star$ gives a differential form $\Omega_0^{\mathrm{ext}}$ such that:
\be
\sigma_0^\star \Omega_0^{\mathrm{ext}} \,=\, \omega \,.
\ee
Nevertheless $(\mathbf{K}^\star,\, \Omega_0^{\mathrm{ext}})$ is not yet the symplectic manifold we were searching for because $\Omega_0^{\mathrm{ext}}$, even if non-degenerate, is, in general, not closed outside $\sigma_0(\mathcal{M})$.
However, as it is proven in \cite[lemma $39.1$ at page $318$]{Guillemin-Sternberg1990-Symplectic_techniques}, a differential form $\alpha$ defined in a tubular neighborhood of $\sigma_0(\mathcal{M})$, say $\tilde{\mathcal{M}}$, such that:
\be
\dd \alpha \,=\, -\dd \Omega_0^{\mathrm{ext}} \;\; \qquad \text{and} \qquad \;\; \alpha\bigr|_{\sigma_0(\mathcal{M})} \,=\, 0 \,,
\ee
can always be added to $\Omega_0^{\mathrm{ext}}$.
Because of the properties of $\alpha$, the manifold $(\tilde{\mathcal{M}},\, \Omega)$, where $\Omega \,=\, \Omega_0^{\mathrm{ext}}\bigr|_{\tilde{\mathcal{M}}} + \alpha$, is the symplectic manifold we were searching for.

The construction of the differential form $\alpha$, as presented in \cite{Guillemin-Sternberg1990-Symplectic_techniques}, requires the choice of a retraction of $\mathbf{K}^\star$ into $\mathcal{M}$ and to construct its flow.
This may be not easy from the computational point of view.
For this reason, in the next section we propose a way of constructing $\Omega$ related with the choice of a connection on the bundle $\mathcal{M} \to \mathcal{M}/K$ ($K$ denoting the distribution associated with $K_m$) which, even if is less general, is more practical from the computational point of view and that also allows for studying the possibility of defining a Poisson structure on the original (pre-symplectic) manifold $\mathcal{M}$.


\subsubsection{The coisotropic embedding theorem via connections}
\label{Subsubsec:The coisotropic embedding theorem via connections}

The splitting $\mathbf{T}_m\mathcal{M} \,=\,\ W_m \oplus K_m$ considered in the previous section can be given in terms of a connection on the bundle $\mathcal{M} \to \mathcal{M} / K$.
Indeed, such a connection is specified by selecting a $1-1$ tensor field on $\mathcal{M}$:
\be 
P \;\; : \;\; \mathfrak{X}(\mathcal{M}) \to \mathfrak{X}^v(\mathcal{M}) \,,
\ee
which is an idempotent operator onto the space of vertical vector fields on $\mathcal{M}$ and which is invariant with respect to the flow of the elements in $\mathfrak{X}^v(\mathcal{M})$.
Given a basis for the module $\mathfrak{X}(\mathcal{M})$, say $\{\,V_j\,\}_{j=1,...,\mathrm{dim}\mathcal{M}}$\footnote{Note that $\mathrm{dim}\mathcal{M}$ may be infinite and, thus, $j$ may should be considered to be an infinite-dimensional label.}, the projector $P$ can be written as:
\be
P \,=\, P^j \otimes V_j \,,
\ee
for some $1$-forms $P^j$ on $\mathcal{M}$.
Denoting by $\{\,\mu_j\,\}_{j=1,...,\mathrm{dim}K^\star}$ a system of coordinates on $K^\star_m$, the term $\omega_{\mathbf{K}\oplus \mathbf{K}^\star}\circ \mathrm{pr}$ reads:
\be
\omega_{\mathbf{K}\oplus \mathbf{K}^\star}\circ \mathrm{pr} \,=\, \dd \mu_j \wedge P^j \,.
\ee
With this in mind, the structure $\Omega_0^{\mathrm{ext}}$ on $\mathbf{K}^\star$ reads:
\be
\Omega_0^{\mathrm{ext}} \,=\, \tau^\star \omega + \dd \mu_j \wedge P^j  \,.
\ee
This structure is not closed because:
\be
\dd \mu_j \wedge P^j \,=\, \dd (\mu_j P^j) - \mu_j \dd P^j \,.
\ee
Therefore, the term which is missing to make $\Omega_0^{\mathrm{ext}}$ closed is exactly the term:
\be
\alpha \,=\, \mu_j \dd P^j  \,,
\ee
which has the properties of the differential form $\alpha$ of the previous section since it vanishes on $\sigma_0(\mathcal{M})$ where $\mu_j = 0$.

Therefore, the symplectic manifold obtained via the coisotropic embedding theorem is any tubular neighborhood of $\sigma_0(\mathcal{M})$ into $\mathbf{K}^\star$ equipped with the symplectic structure:
\be
\Omega \,=\, \tau^\star \omega + \dd \mu_j \wedge P^j + \mu_j \dd P^j \,.
\ee

\subsection{Poisson brackets on pre-symplectic manifolds}
\label{Subsec:Poisson brackets on pre-symplectic manifolds}

Now, we are going to see how, depending on the properties of the connection chosen, the coisotropic embedding theorem can be used to construct Poisson brackets.
In particular we will see that if the connection is closed or has zero curvature, a Poisson bracket can be directly defined on the original pre-symplectic manifold.
On the other hand, if the connection has non-zero curvature, a Poisson bracket can be defined only on the enlarged manifold $\tilde{\mathcal{M}}$.

\subsubsection{The closed case}
\label{Subsubsec:The closed case}
Let us focus for a moment on the case where $\dd P^j = 0$ and let us see how a Poisson structure can be induced on the original pre-symplectic manifold $\mathcal{M}$ starting from $\Omega$.

The symplectic structure $\Omega$, at each point $\tilde{m}$ of $\tilde{M}$, is the sum of a non-degenerate structure having components only on $W_m$ and a non-degenerate structure having components only on $K_m \oplus K^\star_m$:
\be
\begin{split}
\Omega_{\tilde{m}}(\tilde{X}_W + \tilde{X}_K + \tilde{X}_{K^\star},\, \cdot\,) \,&=\, \tau^\star \omega_{\tilde{m}}(\tilde{X}_W, \,\cdot\,) + \dd \mu_j \wedge P^j_{\tilde{m}} (\tilde{X}_K + \tilde{X}_{K^\star},\, \cdot \,) \,=:\\
\;&=:\, \Omega_{W_m}(\tilde{X}_W, \,\cdot\,) + \Omega_{K_m \oplus K_m^\star}(\tilde{X}_K + \tilde{X}_{K^\star},\,\cdot\,)  
\end{split}
\ee
where $\tilde{X}_W \in W_m$, $\tilde{X}_K \in K_m \,\, \forall \, \tilde{m} \in \tilde{\mathcal{M}}$ and $\tilde{X}_{K^\star} \in K^\star_m \,\, \forall \, \tilde{m} \in \tilde{\mathcal{M}}$.
This means that $\Omega$ results as the direct sum of two closed forms:
\be \label{Eq:decomposition omega}
\Omega \,=\, \Omega_W \oplus \Omega_{K \oplus K^\star} 
\ee 
which, restricted respectively to the distributions $W$ and $K \oplus K^\star$, are non-degenerate.
Consequently, the Poisson bi-vector field on $\tilde{\mathcal{M}}$, say $\Lambda$, associated with the symplectic structure $\Omega$, reads:
\be 
\Lambda \,=\, \Lambda_W \oplus \Lambda_{K \oplus K^\star}
\ee
where $\Lambda_W$ is a Poisson bi-vector field belonging, at each $\tilde{m} \in \tilde{\mathcal{M}}$, to $W_m \wedge W_m$ whereas $\Lambda_{K \oplus K^\star}$ is a Poisson bi-vector field belonging, at each $\tilde{m} \in \tilde{\mathcal{M}}$ to $K_m \oplus K_m^\star \wedge K_m \oplus K_m^\star$.
Now, since $\Lambda$ is the Poisson bi-vector field associated with a symplectic structure, it satisfies the following:
\be
[\Lambda,\, \Lambda]_S \,=\, 0
\ee
where $[\,\cdot\,,\,\cdot\,]_S$ denote  the Schouten-Njienhuis brackets, which is equivalent to the fact that the bracket associated to $\Lambda$, $\{f,\,g\} \,=\, \Lambda(\dd f,\, \dd g)$, satisfies the Jacobi identity.
Now, since $\Lambda_W$ is the inverse (restricted to $W$) of the closed (and non-degenerate when restricted to $W$) form $\Omega_W$, it satisfies: 
\be \label{Eq:schouten lambdaW}
\left[\Lambda_W,\, \Lambda_W \right]_S \,=\,0 
\ee
itself.
Another way of proving this latter equality is the following.
The Schouten bracket of $\Lambda$ with itself reads:
\be
[\Lambda,\, \Lambda]_S \,=\, [\Lambda_W,\, \Lambda_W]_S + 2 [\Lambda_W,\, \Lambda_{K \oplus K^\star}]_S + [\Lambda_{K \oplus K^\star},\, \Lambda_{K \oplus K^\star}] 
\ee
where the second term on the right hand side vanishes because, being $P$ defined on $\mathcal{M}$, $W$ is invariant with respect to elements in $K^\star$ and, being $P$ a connection, $W$ is invariant with respect to $K$, i. e., $W$ commutes both with $K^\star$ and with $K$.
What is more, the last term on the right hand side vanishes because: 
\be
\Lambda_{K \oplus K^\star} \,=\, \frac{\de}{\de \mu_j} \wedge V_j 
\ee
and:
\be
\left[\frac{\de}{\de \mu_j},\, V_k\right] \,=\, 0 \,.
\ee
With this in mind \eqref{Eq:schouten lambdaW} is a straightforward consequence. \\
Now, the bi-vector field $\Lambda_W$ satisfying \eqref{Eq:schouten lambdaW} can be used to define a Poisson bracket on the pre-symplectic manifold $\mathcal{M}$.
Indeed, since $W$ is a distribution on $\mathcal{M}$, seen as a distribution on $\tilde{M}$ it clearly commutes with the distribution $K^\star$ generated by $\frac{\de}{\de \mu_j}$.
This means that: 
\be
\left[\Lambda_W,\, \frac{\de}{\de \mu_j}\right]_S \,=\,0 \;\;\; \forall \,\, j=1,..., \mathrm{dim}K^\star \,.
\ee
Consequently, the bi-vector field $\Lambda_W$ is projectable onto $\mathcal{M}$ via $\tau \;:\; \tilde{\mathcal{M}} \to \mathcal{M}$ to the following bi-vector field:
\be 
\lambda_W \,=\, \tau_\star \Lambda_W \,\, \in \bigwedge^2(\mathcal{M}) \,.
\ee
The latter also has a vanishing Schouten bracket with itself because of the following equalities:
\be
[\lambda_W,\, \lambda_W]_S \,=\, [\tau_\star \Lambda_W,\, \tau_\star \Lambda_W]_S \,=\, \tau_\star \underbrace{[\Lambda_W,\, \Lambda_W]}_{=0} \,=\, 0
\ee
and, thus, it defines a Poisson bracket on $\mathcal{M}$ in the following way:
\be \label{Eq:poisson bracket M}
\{\, f,\,g \,\} \,=\, \lambda_W(\dd f,\, \dd g) 
\ee
for $f,\, g \in \mathcal{F}(\mathcal{M})$.

\subsubsection{Zero-curvature case}
\label{Subsubsec:Zero-curvature case}

The previous construction can be extended to the case where $P^j$ is not closed but only horizontally-closed, i. e., to the case where:
\be
\dd_H P^j \,=\, \dd P^j \left(\,(\mathbb{1} - P)(\,\cdot\,),\, (\mathbb{1}- P)(\,\cdot\,)\, \right) \,=\, 0 \,.
\ee
Indeed, when $P^j$ is not closed, the structure $\Omega$ reads:
\be
\Omega \,=\, \Omega_W \oplus \Omega_{K \oplus K^\star} \oplus \alpha
\ee
where $\Omega_W$ and $\Omega_{K \oplus K^\star}$ are closed and, in general, $\alpha \,=\, \mu_j \dd P^j$ has components both on $W$ and on $K \oplus K^\star$, i.e.:
\be
\begin{split}
\alpha_{\tilde{m}}(\tilde{X}_W + \tilde{X}_K + \tilde{X}_{K^\star},\,\cdot\,) \,&=\, \alpha_{\tilde{m}}(\tilde{X}_W,\,\cdot\,) + \alpha_{\tilde{m}}(\tilde{X}_K + \tilde{X}_{K^\star},\,\cdot\,) \,=:\, \\ 
\,&=\,{\alpha_W}_{\tilde{m}} (\tilde{X}_W,\,\cdot\,) + {\alpha_{K\oplus K^\star}}_{\tilde{m}}(\tilde{X}_K + \tilde{X}_{K^\star} , \, \cdot \,) \,=\, \\
\,&=\, \mu_j \dd_H P^j_{\tilde{m}}(\tilde{X}_W,\,\cdot\,) + \mu_j \dd_V P^j (\tilde{X}_K + \tilde{X}_{K^\star},\, \cdot \,) \,,
\end{split}
\ee
where $\dd_V P^j (X,\, Y) \,=\, \dd P^j(P(X),\, P(Y))$ and $\dd_H P^j (X,\, Y) \,=\, \dd P^j((\mathbb{1}-P)(X),\, (\mathbb{1}-P)(Y) )$. \\
Now, if $\dd_H P^j \,=\, 0$, i. e., if $P$ has zero curvature, $\Omega$ reads:
\be
\Omega \,=\, \Omega_W \oplus \tilde{\Omega}_{K \oplus K^\star}
\ee
where $\tilde{\Omega}_{K \oplus K^\star} \,=\, \Omega_{K \oplus K^\star} + \alpha_{K \oplus K^\star}$.
Therefore, again $\Lambda$ reads:
\be
\Lambda \,=\, \Lambda_W \oplus \tilde{\Lambda}_{K \oplus K^\star}
\ee
with $\Lambda_W$ satisfying:
\be
\left[\Lambda_W,\, \Lambda_W \right] \,=\, 0
\ee
since it comes from a closed $2$-form $\Omega_W$.
Therefore, also in this case the previous construction can be performed, giving rise to the Poisson bracket \eqref{Eq:poisson bracket M} on $\mathcal{M}$.

\subsubsection{The non-closed case}
\label{Subsubsec:The non-closed case}

The general case where neither $\dd_H P^j = 0$ nor $\dd_V P^j \,=\, 0$\footnote{Recall that $\dd_V P^j \,=\, \dd P^j (P(\,\cdot\,),\, P(\,\cdot\,))$.} does not allow to directly define a Poisson bracket on $\mathcal{M}$ using the Poisson bracket defined on $\tilde{\mathcal{M}}$.

Indeed, in that case the structure $\Omega$ reads:
\be
\Omega \,=\, \tilde{\Omega}_W \oplus \tilde{\Omega}_{K \oplus K^\star} \,,
\ee
where:
\be
\tilde{\Omega}_W \,=\, \Omega_W \oplus \alpha_W \,, \qquad \tilde{\Omega}_{K \oplus K^\star} \,=\, \Omega_{K \oplus K^\star} \oplus \alpha_{K \oplus K^\star} \,,
\ee
with:
\be
\alpha_W \,=\, \mu_j \dd_H P^j \,, \qquad \alpha_{K \oplus K^\star} \,=\, \mu_j \dd_V P^j \,. 
\ee
However, in this case even if $\tilde{\Omega}_W$ and $\tilde{\Omega}_{K \oplus K^\star}$ are non-degenerate when restricted to $W$ and $K \oplus K^\star$ respectively, they are not closed.
Therefore, the corresponding bi-vector fields $\tilde{\Lambda}_W$ and $\tilde{\Lambda}_{K \oplus K^\star}$ such that:
\be
\Lambda \,=\, \tilde{\Lambda}_W \oplus \tilde{\Lambda}_{K \oplus K^\star} \,,
\ee
do not satisfy:
\be
\left[\tilde{\Lambda}_W,\, \tilde{\Lambda}_W \right]_S \,=\, 0 \,, \qquad \, \left[ \tilde{\Lambda}_{K \oplus K^\star},\, \tilde{\Lambda}_{K \oplus K^\star} \right]_S \,=\,0 \,,
\ee 
and, thus, $\tilde{\Lambda}_W$ can not define a Poisson bracket on $\mathcal{M}$.

The only natural construction in this case seems to be to consider the Poisson bracket associated with $\Omega$ on the whole $\tilde{\mathcal{M}}$ restricted to the subalgebra of functions on $\tilde{\mathcal{M}}$ being pull-back (via $\tau$) of functions on $\mathcal{M}$.
That is, one can consider two functions $\tilde{f},\, \tilde{g} \in \mathcal{F}(\tilde{\mathcal{M}})$ such that $\tilde{f} \,=\, \tau^\star f$ and $\tilde{g} \,=\, \tau^\star g$ with $f,\, g \in \mathcal{F}(\mathcal{M})$ and to consider their bracket computed with respect to $\Lambda$
\be
\{\tilde{f},\, \tilde{g}\} \,=\,\Lambda(\dd \tilde{f},\, \dd \tilde{g}) \,=\, \tilde{\Lambda}_W (\dd \tilde{f},\, \dd \tilde{g}) \,.
\ee
Even if, due to the fact that $\Lambda$ is a bi-vector field coming from a symplectic structure, this bracket satisfies the Jacobi identity, it can not be used to induce a bracket on $\mathcal{M}$ because, in general, since the term $\alpha_{W}$ added to $\Omega_W$ contains a dependence on the variables $\mu_j$, the function $\{\tilde{f},\, \tilde{g}\}$ is not the pull-back of a function on $\mathcal{M}$ as well.
Indeed, in general it will be of the form
\be 
\{\tilde{f},\, \tilde{g}\} \,=\, \tilde{h} + H
\ee 
where $\tilde{h} \,=\, \tau^\star h$ with $h \in \mathcal{F}(\mathcal{M})$ and $H \in \mathcal{F}(\tilde{\mathcal{M}})$.

In the next sections we will see how all these constructions can be used to define a Poisson bracket on the solution space of non-Abelian gauge theories.
For this reason, even if it is already extensively described in the companion paper \cite{Ciaglia-DC-Ibort-Marmo-Schiav-Zamp2021-Cov_brackets_toappear}, we first devote Sect. \ref{Sec:Multisymplectic formulation of Classical Field Theories} to recall the multisymplectic formulation of first order Hamiltonian field theories and how in this formalism the solution space of the theory is canonically a pre-symplectic manifold (in particular, it is genuinely pre-symplectic within gauge theories).
Then, in Sect. \ref{Sec:Covariant Poisson bracket on the solution space of Yang-Mills theories} we will apply the construction of the present section to the solution space of Yang-Mills theories.


\section{Multisymplectic formulation of Classical Field Theories}
\label{Sec:Multisymplectic formulation of Classical Field Theories}

In this section we recall the multisymplectic formulation of first order Hamiltonian field theories for which we refer to \cite{Ibort-Spivak2017-Covariant_Hamiltonian_YangMills, Ciaglia-DC-Ibort-Marmo-Schiav-Zamp-2022-Symmetry, Ciaglia-DC-Ibort-Marmo-Schiav-Zamp2021-Cov_brackets_toappear} and references therein.
Since it is already extensively discussed in the companion paper \cite{Ciaglia-DC-Ibort-Marmo-Schiav-Zamp2021-Cov_brackets_toappear}, here, for the sake of completeness, we recall the main features of the formalism but we will avoid all the proofs.

Sect. \ref{Subsec:Carrier space: the Covariant Phase Space} is devoted to recall the construction of the carrier space where one settles the multisymplectic description of first order Hamiltonian field theories.
In Sect. \ref{Subsec:The space of dynamical fields and the Schwinger-Weiss variational principle} we recall how to formulate the variational principle à là Schwinger-Weiss.
Finally, in Sect. \ref{Subsec:The canonical pre-symplectic structure and the canonical formalism near the boundary} we recall how a canonical $2$-form on the solution space of the theory emerges.


\subsection{Carrier space: the Covariant Phase Space}
\label{Subsec:Carrier space: the Covariant Phase Space}

A multisymplectic formulation of a classical field theory starts with:
\begin{itemize}
\item A reference space-time, i.e., an $n$-dimensional ($n=d+1$) smooth differential manifold $\m$, eventually with boundary $\de \m$, where we use a system of local coordinates $(\, x^\mu \,) ,\, \mu=0,...,d$ defined on an open set $U_\m \subset \m$.
Actually, it is not required for $\m$ to be a space-time in the sense that it carries a Lorentzian metric, but rather it is only required to be orientable, i.e., to carry a volume form, say $vol_\m \in \Omega^n(\m)$;
\item A fibre bundle over $\m$, say $\pi \;\; : \;\; \mathbb{E} \to \m$, whose typical fibre is denoted by $\mathcal{E}$ and where we use the system of local fibred coordinates $(\, x^\mu,\, u^a \,),\, \mu=0,...,d,\, a=1,...,r$ defined on an open set $U_{\mathbb{E}} \subset \mathbb{E}$.
Sections of $\pi$ are local functions on $\m$ with values in $\mathcal{E}$, they are denoted by $\phi$ and represent the \textit{configuration fields} of the theory.
In the system of local fibred coordinates chosen they read:
\be 
\phi \;\; :\;\; U_\m \to U_{\mathbb{E}} \;\; :\;\; x^\mu \mapsto \phi^a(x) \,.
\ee
The carrier space where we settle the description of the field theory is the reduced dual of the first order jet bundle $\mathbf{J}^1\pi$ (see \cite{Ciaglia-DC-Ibort-Marmo-Schiav-Zamp2021-Cov_brackets_toappear}), which we call \textit{Covariant Phase Space}\footnote{Note that within the literature sometimes people refer to Covariant Phase Space as the solution space of the equations of motion, whereas for us it will be called simply \textit{solution space} or \textit{Euler-Lagrange space}.} and we denote by $\pe$.
As in \cite{Ciaglia-DC-Ibort-Marmo-Schiav-Zamp2021-Cov_brackets_toappear}, we denote by $(x^\mu,\, u^a,\, \rho^\mu_a),\, \mu=0,...,d,\, a=1,...,r$, a system of adapted fibred coordinates on $\pe$ defined on an open set $U_{\pe} \subset \pe$. 
\item A \textit{Hamiltonian}, i.e., a local section of the projection $\kappa$ appearing in $Eq. 5$ of \cite{Ciaglia-DC-Ibort-Marmo-Schiav-Zamp2021-Cov_brackets_toappear}:
\be
\mathrm{H} \;\; :\;\; U_{\pe} \to U_{\mathbf{J}^\dag \pi} \;\; :\;\; (x^\mu,\, u^a,\, \rho^\mu_a) \mapsto (x^\mu,\, u^a,\, \rho^\mu_a,\, \rho_0 \,=\, H(x,\, u,\, \rho)) \,,
\ee
which defines the local function $H$\footnote{In the rest of the paper we will refer to \textit{Hamiltonian} equivalently to indicate the section $\mathrm{H}$ and the function $H$}.
Now, the generic $1$-semibasic $n$-form on $\mathbb{E}$ reads (see \cite{Ciaglia-DC-Ibort-Marmo-Schiav-Zamp2021-Cov_brackets_toappear}):
\be
\mathbf{w} \,=\, \rho^\mu_a \dd u^a \wedge i_{\de_\mu} vol_\m + \rho_0 vol_\m \,.
\ee
The choice of a Hamiltonian defines a unique differential $n$-form on $\pe$ which is the pull-back via $\mathrm{H}$ of $\mathbf{w}$\footnote{The minus sign is a matter of convention.}:
\be
\Theta_H \,=\, (-\mathrm{H})^\star \mathbf{w} \,=\, \rho^\mu_a \dd u^a \wedge i_{\de_\mu} vol_\m - H(x,\,u,\, \rho) vol_\m \,.
\ee
Its differential is denoted by $-\dd \Theta_H$ and is a \textit{multisymplectic form}\footnote{We recall that a multisymplectic form is a closed non-degenerate $n$-form, where by non-degenerate we mean that $i_X \dd \Theta_H \,=\,0 \,\, \implies \,\, X = 0$.}, from which the name of the formalism we are using.
\end{itemize}

\subsection{The space of dynamical fields and the Schwinger-Weiss variational principle}
\label{Subsec:The space of dynamical fields and the Schwinger-Weiss variational principle}

The dynamical content of the theory will be implemented by means of a variational principle à là Schwinger-Weiss.
The first ingredient to formulate it, is an action functional to extremize.
With the ingredients introduced in the previous section in hand, an action functional can be defined in the following way:
\be \label{Eq:action functional}
\ac_\chi \,=\, \int_\m \chi^\star \Theta_H \,=\, \int_\m \left[\, P^\mu_a \de_\mu \phi^a - H(\chi) \,\right] vol_\m \,,
\ee
where $\chi$ is a section of $\delta_1$.
In particular we chose $\chi$ to be a section of the type introduced in $Def. 2.1$ of \cite{Ciaglia-DC-Ibort-Marmo-Schiav-Zamp2021-Cov_brackets_toappear}, i.e., a section of $\delta_1$ which is the composition of a section $\phi$ of $\pi$ and a section $P$ of $\delta^1_0$, $\chi \,=\, P \circ \phi \,=:\,(\phi,\, P)$.

Note that $\ac$ is a real-valued function on $\Gamma_{\textsc{split}}(\delta_1)$.
$\Gamma_{\textsc{split}}(\delta_1)$, being (a subset of) a space of smooth sections, is a Frechet space.
However, in order to avoid the amount of technicalities needed to define a Cartan-like differential calculus on Frechet spaces\footnote{For which we refer to \cite{Michor-Kriegl1997-Convenient_setting} and references therein.}, it would be desirable for $\ac$ to be defined actually on a Banach manifold.
The way we will address this problem is to extend (by continuity) $\ac$ to the closure of $\Gamma_{\textsc{split}}$ with respect to a Banach norm in which $\ac$ is continuous.
The existence of such a norm is clearly not guaranteed in general and it depends on the particular $\ac$ one has in hand.
However, from now on we assume it to exist in general and we prove that this is true case by case in the examples considered.
We denote the closure of $\Gamma_{\textsc{split}}(\delta_1)$ by $\fpe$ and we still denote by $\chi \,=\, (\phi,\, P)$ elements of $\fpe$.
The induced space of $\phi$'s will be denoted by $\mathcal{F}_{\mathbb{E}}$.
We will refer to $\fe$ as the \textit{space of configuration fields} and to $\fpe$ as the \textit{space of dynamical fields} of the theory.

Now, given a tangent vector to $\fpe$ at $\chi$, say $\mathbb{X}_\chi$, the variation of $\ac$ in the direction of $\mathbb{X}_\chi$ reads \cite{Ibort-Spivak2017-Covariant_Hamiltonian_YangMills, Ciaglia-DC-Ibort-Marmo-Schiav-Zamp2021-Cov_brackets_toappear}:
\be \label{Eq:first variational formula integral}
\delta_{\mathbb{X}_\chi} \ac_\chi \,=\, \int_\m \chi^\star \left[\, i_X \dd \Theta_H \,\right] + \int_{\de \m} \chi_{\de \m}^\star \left[\, i_X\Theta_H \,\right] \,,
\ee
where $X$ is a vector field over $\pe$ defined on an open neighborhood of the image of $\chi$ which is an extension of $\mathbb{X}_\chi$\footnote{It means that $X$ agree with $\mathbb{X}_\chi$ when evaluated on the image of $\chi$. 
Note that since the argument of the integral is evaluated along the image of $\chi$, this term does not depend on the particular extension chosen.} and $\chi_{\de \m} \,=\, \chi\bigr|_{\de \m}$ is the restriction of $\chi$ to $\de \m$.
The terms appearing in \eqref{Eq:first variational formula integral}, being all linear in $\mathbb{X}_\chi$ can be interpreted in terms of differential forms on $\fpe$ as explained in \cite{Ibort-Spivak2017-Covariant_Hamiltonian_YangMills, Ciaglia-DC-Ibort-Marmo-Schiav-Zamp2021-Cov_brackets_toappear}, and \eqref{Eq:first variational formula integral} can be written as:
\be
\dd \ac \,=\, \mathbb{EL} + \Pi_{\de \m}^\star \alpha^{\de \m}
\ee
where $\mathbb{EL} \in \Omega^1(\fpe)$ is defined by:
\be
\mathbb{EL}_\chi(\mathbb{X}_\chi) \,=\, \int_\m \chi^\star \left[\, i_X \dd \Theta_H \,\right] \,,
\ee
the map $\Pi_{\de \m}$ is the restriction map to the boundary:
\be
\Pi_{\de \m} \;\; : \;\; \fpe \to \fpe^{\de \m} \;\; :\;\; \chi \mapsto \chi\bigr|_{\de \m} \,, 
\ee
($\fpe^{\de \m}$ being the space of restrictions of $\chi$'s to $\de \m$) and $\Pi^\star_{\de \m} \alpha^{\de \m} \in \Omega^1(\fpe)$ is defined by:
\be
\Pi_{\de \m}^\star \alpha^{\de \m}_{\chi}(\mathbb{X}_\chi) \,=\, \int_{\de \m} \chi_{\de \m}^\star \left[\, i_X \Theta_H \, \right] \,,
\ee
where $\alpha^{\de \m} \in \Omega^1(\fpe^{\de \m})$ is defined by:
\be
\alpha^{\de \m}_{\chi_{\de \m}}(\mathbb{X}_{\chi_{\de \m}}) \,=\, \int_{\de \m} \chi_{\de \m}^\star \left[\, i_X \Theta_H \, \right] \,,
\ee
($\mathbb{X}_{\chi_{\de \m}} \in \mathbf{T}_{\chi_{\de \m}}\fpe^{\de \m}$).

Now, the \textit{Schwinger-Weiss variational principle} states that the dynamical fields appearing in nature are those for which the variation (along any direction) of the action functional only depends on boundary terms, that is, on terms only depending on the restrictions of the $\chi$'s to $\de \m$.
Evidently, the term $\Pi^\star_{\de \m} \alpha^{\de \m}_\chi$ is a boundary term and, thus, the dynamical fields appearing in nature are those $\chi \in \fpe$ for which:
\be
\mathbb{EL}_\chi(\mathbb{X}_\chi) \,=\, 0 \;\;\; \forall \,\, \mathbb{X}_\chi \in \mathbf{T}_{\chi} \fpe \,.
\ee
As it is proven in \cite{Ibort-Spivak2017-Covariant_Hamiltonian_YangMills, Ciaglia-DC-Ibort-Marmo-Schiav-Zamp2021-Cov_brackets_toappear}, such $\chi$'s satisfy:
\be
\chi^\star \left[\, i_X \dd \Theta_H \,\right] \,=\, 0 \;\;\; \forall \,\, X \in \mathfrak{X}(U^{(\chi)})\,,
\ee
where $U^{(\chi)}$ is an open neighborhood of the image of $\chi$.
The latter equation, in the system of local coordinates chosen on $\pe$ reads:
\be
\begin{cases}
\frac{\de \phi^a}{\de x^\mu} \,=\, \frac{\de H}{\de \rho^\mu_a}\biggr|_{\chi} \,, \\
\frac{\de P^\mu_a}{\de x^\mu} \,=\, -\frac{\de H}{\de u^a}\biggr|_{\chi} 
\end{cases}
\ee
which are the so-called \textit{De Donder-Weyl equations} also called sometimes \textit{covariant Hamilton equations}.

The space of zeroes of $\mathbb{EL}$, i.e., the space of solutions of the De Donder-Weyl equations is what we will refer to as \textit{the solution space} throughout the whole paper and we will denote by:
\be
\elag_\m \,:=\, \left\{\, \chi \in \fpe \;\; :\;\; \mathbb{EL}_\chi \,=\, 0 \,\right\} \,.
\ee
We will assume it to be a smooth immersed submanifold of $\fpe$.


\subsection{The canonical pre-symplectic structure and the canonical formalism near the boundary}
\label{Subsec:The canonical pre-symplectic structure and the canonical formalism near the boundary}

As it is extensively discussed in \cite{Ciaglia-DC-Ibort-Marmo-Schiav-Zamp2021-Cov_brackets_toappear}, the space $\elag_\m$ carries a canonical $2$-form that can be used to define, in some cases, a Poisson bracket on $\elag_\m$.
It is related with the differential $1$-form $\alpha^{\de \m}$ of Sect. \ref{Subsec:The space of dynamical fields and the Schwinger-Weiss variational principle} in the following way.

The differential of $\Pi_{\de \m}^\star\alpha^{\de \m}$ is a $2$-form on $\fpe$ given by\footnote{The minus sign is a matter of convention.} \cite{Ciaglia-DC-Ibort-Marmo-Schiav-Zamp2021-Cov_brackets_toappear}:
\be \label{Eq:definition canonical structure}
-\dd \Pi_{\de \m}^\star \alpha^{\de \m}_\chi (\mathbb{X}_\chi,\, \mathbb{Y}_\chi) \,=:\, \Pi^\star_{\de \m} \Omega^{\de \m}(\mathbb{X}_\chi,\, \mathbb{Y}_\chi) \,=\, \int_{\de \m} \left[ i_X i_Y \dd \Theta_H \right] \,, 
\ee
where $X$ and $Y$ are any two extensions of $\mathbb{X}_\chi$ and $\mathbb{Y}_\chi$ to vector fields defined in an open neighborhood of the image of $\chi$.
It can be equivalently defined for any slice\footnote{By \textit{slice} we mean a co-dimension one hypersurface of $\m$ which split the space-time $\m$ in two regions, say $\m_+$ and $\m_-$.} $\Sigma$ of $\m$ by formally substituting $\de \m$ with $\Sigma$:
\be \label{Eq:Hamiltonian functional}
\pssos_\chi(\mathbb{X}_\chi,\, \mathbb{Y}_\chi)\,:=\, \int_\Sigma \left[ i_X i_Y \dd \Theta_H \right] \,.
\ee
Now, in \cite{Ciaglia-DC-Ibort-Marmo-Schiav-Zamp2021-Cov_brackets_toappear} it is proven that $\pssos$ does not depend on the particular $\Sigma$ chosen if evaluated on $\chi \in \elag_\m$.

In \cite{Ibort-Spivak2017-Covariant_Hamiltonian_YangMills, Ciaglia-DC-Ibort-Marmo-Schiav-Zamp2021-Cov_brackets_toappear} it is also proved that locally $\elag_\m$ is isomorphic with the space of solutions of the pre-symplectic Hamiltonian system $(\fpe^\Sigma,\, \Omega^\Sigma,\, \mathcal{H})$, where $\fpe^{\Sigma}$ is the space of fields restricted to $\Sigma$, $\os$ is the $2$-form on $\fpe^\Sigma$ appearing in  \eqref{Eq:definition canonical structure} and $\mathcal{H}$ is a function on $\fpe^{\Sigma}$ which, by fixing a coordinate system on $\m$ such that $\Sigma$ is the hypersurface with $x^0 \,=\, x^0_\Sigma$\footnote{This can be done without loss of generality by virtue of the embedding theorem being $\Sigma$ a smooth submanifold of $\m$.}, reads:
\be
\mathcal{H}(\chi_\Sigma) \,=\, \int_\Sigma \left[\, - \beta^k_a \partial_k \varphi^a + H(\chi_\Sigma)  \,\right] vol_\Sigma \,,
\ee
where $\chi_\Sigma \in \fpe^\Sigma$ has the following coordinate expression:
\be 
\chi_\Sigma(\underline{x}) \,=\, \left(\, \phi^a\bigr|_\Sigma (\underline{x}) ,\, P^0_a\bigr|_\Sigma (\underline{x}),\, P^k_a\bigr|_\Sigma (\underline{x}) \,\right) \,=:\, \left(\, \varphi^a (\underline{x}),\, p_a (\underline{x}), \beta^k_a (\underline{x}) \,\right) \,,
\ee
where $\underline{x} \in \Sigma$ and where we have distinguished the component $P^0_a$ which, in the system of coordinate chosen, is transversal to $\Sigma$ from the tangent components $P^k_a$.
By means of "solutions of the pre-symplectic system" we mean the integral curves of the vector field $\mathbb{\Gamma}$ on $\fpe^\Sigma$ which is solution of the canonical equation:
\be
i_{\mathbb{\Gamma}} \os \,=\, \dd \mathcal{H} \,.
\ee
It is well known \cite{Gotay-Nester-Hinds1978-DiracBergmann_constraints} that, being $\os$ pre-symplectic, the latter equation can not be considered as it is as a well-posed equation for $\mathbb{\Gamma}$ but rather one should apply the so-called \textit{pre-symplectic constraint algorithm} (PCA).
Such algorithm (that we recalled in Appendix $A1$ of \cite{Ciaglia-DC-Ibort-Marmo-Schiav-Zamp2021-Cov_brackets_toappear}) allows to find, eventually, a smooth immersed submanifold of $\fpe^\Sigma$ that we will denote by $\mathfrak{i}_\infty \;\; :\;\; \mathcal{M}_\infty \hookrightarrow \fpe^\Sigma$, on which the canonical equation:
\be 
i_{\mathbb{\Gamma}_\infty} \os_\infty \,=\, \dd \mathcal{H}_\infty
\ee 
is well posed, where $\os_\infty \,=\, \mathfrak{i}_\infty^\star \os$, $\mathcal{H}_\infty \,=\, \mathfrak{i}_\infty^\star \mathcal{H}$ and $\mathbb{\Gamma}_\infty \in \mathfrak{X}(\mathcal{M}_\infty)$.
The integral curves of such a $\mathbb{\Gamma}_\infty$ immersed into $\fpe^\Sigma$ via $\mathfrak{i}_\infty$ are the solutions of the original pre-symplectic system considered.

In \cite{Ibort-Spivak2017-Covariant_Hamiltonian_YangMills, Ciaglia-DC-Ibort-Marmo-Schiav-Zamp2021-Cov_brackets_toappear} it is proved that such solutions are at least locally in one-to-one correspondence with extrema of $\ac$.
In \cite{Ciaglia-DC-Ibort-Marmo-Schiav-Zamp2021-Cov_brackets_toappear} it is also proved that when $\os_\infty$ is symplectic it gives a Poisson bracket which is equivalent to the one defined by the canonical structure $\pssos$ in the sense that they are related via a diffeomorphism.

Therefore, the idea here is to use again the structure $\os_\infty$ emerging from the PCA applied to Yang-Mills theories to define a Poisson bracket on $\elag_\m$.

The main problem here, is that in this case $\os_\infty$ turns out to be pre-symplectic and, therefore, it does not give rise directly to a Poisson bracket.
Regarding this problem, the authors already started to analyse the case in which $\os_\infty$ is still pre-symplectic in \cite{Ciaglia-DC-Ibort-Marmo-Schiav-Zamp2021-Cov_brackets_toappear} where the first example of gauge theory, that is, free Electrodynamics is considered.
There, the difficulty of defining a Poisson bracket in the pre-symplectic case was overcome by the aid of a flat connection by using an idea developed in \cite{Dubrovin-Giord-Marmo-Sim1993-Poisson_presymplectic}.
However, as we will see, in this example a flat connection can not be chosen and we will use the coisotropic embedding theorem as a tool to define a Poisson bracket on a suitable enlargement of $\elag_\m$.


\section{Covariant Poisson bracket on the solution space of Yang-Mills theories}
\label{Sec:Covariant Poisson bracket on the solution space of Yang-Mills theories}

\subsection{A preliminar example: the magnetic monopole}
\label{Subsec:A preliminar example: the magnetic monopole}

Having in mind the Lagrangian description of the magnetic monopole given in \cite{Bala-Marmo1983-Gauge_theories}, we propose the following multisymplectic Hamiltonian description.

We consider the one-dimensional space-time $\m \,=\, \mathbb{I}$ being an interval of the real line with (global) coordinate $\{ t \}$.
The bundle $\mathbb{E}$ underlying the theory is:
\be
\pi \;\; :\;\; \mathbb{I} \times \mathbb{R}_+ \times SU(2) \to \mathbb{R} \,,
\ee
on which we will use the following set of local coordinates $\{\, t,\, r,\, s \,\}$ where $r$ is a (global) coordinate on $\mathbb{R}_+$ and $\{\,s \,\}$ is a system of local coordinates on $SU(2)$.
Having in mind the description given in \cite{Bala-Marmo1983-Gauge_theories}, we will not consider the whole Covariant Phase space $\pe \,=\, \mathbb{I} \times \mathbf{T}^\star \mathbb{R}_+ \times \mathbf{T}^\star SU(2)$ with local coordinates $\{\, t,\, r, p_r,\, s,\, \alpha_j \,\}_{j=1,2,3}$\footnote{$\alpha_j$ denote the coordinates on the fibres of $\mathbf{T}^\star SU(2)$ associated with the choice of the left-invariant one-forms as a basis for the one-forms on $SU(2)$.}, but rather a submanifold of it, say $\mathcal{M} \,=\, \mathbb{I} \times \mathbf{T}^\star \mathbb{R}_+ \times \mathcal{N}$, where $\mathcal{N}$ is the submanifold of $\mathbf{T}^\star SU(2)$ given by the condition $\alpha_3 \,=\, n$, for some real constant $n$.
The space of dynamical fields in this case is the space of smooth sections of the bundle $\pi \;\;:\;\; \mathcal{M} \to \mathbb{I}$, which, equipped with the $\mathrm{sup}$-norm, is a Banach manifold.

We consider the following Hamiltonian on $\mathcal{M}$:
\be
H \,=\, \frac{1}{2} \left(\, p_r^2 + \frac{\alpha_1^2 + \alpha_2^2}{r^2} \,\right) \,,
\ee
giving rise to the one-form:
\be
\Theta_H \,=\, p_r \dd r + \alpha_j \theta^j - H \dd t \,,
\ee
where $\theta_j$ are the left-invariant one-forms on $SU(2)$.

In this case a one-codimension hypersurface of $\m$ is just a point, say $\{\overline{t} \}$ in $\mathbb{R}$.
Consequently, the $2$-form $\os$ is the following differential form on the finite-dimensional manifold $\mathbf{T}^\star \mathbb{R}_+ \times \mathcal{N}$:
\be
\os \,=\, \dd p_r \wedge \dd r + \dd \alpha_j \wedge \theta^j + \alpha_j \dd \theta^j \,,
\ee
which, by using the fact that $\dd \theta^j \,=\, \frac{1}{2} \epsilon^j_{\ kl} \theta^k \wedge \theta^l$, reads:
\be
\os \,=\, \dd p \wedge \dd r + (\dd \alpha_1 - \alpha_2 \theta_3) \wedge \theta_1 + (\dd \alpha_2 + \alpha_1 \theta_3) \wedge \theta_2 - n \theta_1 \wedge \theta_2 \,.
\ee
What is more, being $\m$ one-dimensional, the Hamiltonian functional $\mathcal{H}$ coincide with the Hamiltonian function $H$
\be 
\mathcal{H} \,=\, \frac{1}{2} \left( p^2 + \frac{\alpha_1^2 + \alpha_2^2}{r^2} \right) \,.
\ee

Since the manifold $\mathbf{T}^\star \mathbb{R}_+$ does not play any relevant role in the construction of the coisotropic embedding and of the related Poisson bracket, we will neglect it and we will refer to $\mathcal{M}$ as $\mathcal{M} \,=\, \mathcal{N}$ with pre-symplectic structure
\be
\omega \,=\, (\dd \alpha_1 - \alpha_2 \theta_3) \wedge \theta_1 + (\dd \alpha_2 + \alpha_1 \theta_3) \wedge \theta_2 - n \theta_1 \wedge \theta_2 \,,
\ee
with $\theta_j$ being the pull-back of the left-invariant one-forms on $SU(2)$ to $\mathcal{M}$.
The kernel of the latter pre-symplectic structure reads
\be
K_m \,=\, \mathrm{span} \left\{\, \bar{X}_3^\uparrow \,\right\}
\ee
where $X_3$ denotes the third left-invariant vector field on $SU(2)$, $X_3^\uparrow$ denotes its canonical lift to $\mathbf{T}^\star SU(2)$, and $\bar{X}_3^\uparrow$ denotes the projection of the latter vector field to $\mathcal{M}$.\\
In this case the complement $W_m$ of $K_m$ inside $\mathbf{T}_m \mathcal{M}$ is taken in the following way.
Consider the local coordinates on $SU(2)$ given by $\{\, s,\, \beta_j \,\}_{j=1,2,3}$ where $\beta_j$ are the coordinates over the fibres associated with the choice of the right-invariant forms as a basis of differential one-forms.
Consider the basis of vector fields over $\mathbf{T}^\star SU(2)$ given by the canonical lift of the right-invariant vector fields, say $Y_j^\uparrow$, and the vertical vector fields $\frac{\de}{\de \beta_j}$.
Project them to $\mathcal{M}$ to the basis of vector fields $\left\{\, \bar{Y}_j^\uparrow,\, \frac{\de}{\de \beta_j} - \dd \alpha_3\left(\frac{\de}{\de \beta_j}\right) \frac{\de}{\de \alpha_3} \,\right\}$.
Then, subtract their components along $\bar{X}_3^\uparrow$, obtaining
\be
W \,=\, \mathrm{span}\left\{\, Z_j \,=\, \bar{Y}_j^\uparrow - \theta_3(\bar{Y}_j^\uparrow) \bar{X}_3^\uparrow \,, \;\; B_j \,=\, \frac{\de}{\de \beta_j} - \dd \alpha_3\left(\frac{\de}{\de \beta_j}\right) \frac{\de}{\de \alpha_3} \,\right\}_{j=1,2,3} \,.
\ee
This choice of $W$ is the one given by the connection whose projector onto the module of horizontal vector fields reads  
\be
P \,=\, \theta_3 \otimes \bar{X}_3^\uparrow \,.
\ee
To construct the symplectic manifold of the coisotropic embedding theorem we consider the dual of the characteristic bundle of $(\mathcal{M},\, \omega)$, i. e. $\mathbf{K}^\star \,\simeq\, \mathbf{T}^\star SU(2)$ where we chose the system of local coordinates $\{\,s,\, \alpha_1,\, \alpha_2,\, \mu\,\}$.
It project onto $\mathcal{M}$ via the projection
\be
\tau \;\; : \;\; \mathbf{T}^\star SU(2) \to \mathcal{M} \;\; :\;\; (s,\, \alpha_1,\, \alpha_2,\, \mu) \mapsto (s,\, \alpha_1,\, \alpha_2) \,.
\ee
The structure $\Omega_0$ along the zero section of $\tau$ reads
\be
\Omega_0 \,=\, \omega + \dd \mu \wedge \theta_3 \,.
\ee
Its extension over a tubular neighborhood of the zero section reads
\be
\Omega_0^{\mathrm{ext}} \,=\, \tau^\star \omega + \dd \mu \wedge \theta_3
\ee
but it is not closed since $\dd \omega \,=\,0$ but $\dd (\dd \mu \wedge \theta_3) \,=\, 0$.
Indeed
\be
\dd \mu \wedge \theta_3 \,=\, \dd (\mu \theta_3) - \mu \dd \theta_3 \,,
\ee
thus the term $\alpha$ needed to made $\Omega_0^{\mathrm{ext}}$ closed is exactly
\be
\alpha \,=\, \mu \dd \theta_3 \,,
\ee
which, indeed, vanish along the zero section of $\tau$.
Therefore, the symplectic manifold of the coisotropic embedding theorem in this case is $\tilde{\mathcal{M}} \,=\, \mathbf{T}^\star SU(2)$ with the symplectic form
\be
\Omega \,=\, \tau^\star \omega + \dd \mu \wedge \theta_3 + \mu \dd \theta_3 \,. 
\ee
Recalling that $\dd \theta_3 \,=\, -\theta_1 \wedge \theta_2$ we are in the case where $\alpha$ has only "components" along $W$, since it does not contain terms in $\theta_3$.
The term $- \theta_1 \wedge \theta_2$ is indeed the curvature of the connection associated with the one form $\theta_3$ which is non-vanishing.
Therefore
\be
\Omega \,=\, \tilde{\Omega}_W + \Omega_{K\oplus K^\star}
\ee
where $\tilde{\Omega}_W \,=\,  \tau^\star \omega - \mu \theta_1 \wedge \theta_2$.

Note that nor $\tilde{\Omega}_W$ nor $\Omega_{K \oplus K^\star}$ are closed and, thus, $\tilde{\Omega}_W$ can not be used to induce a Poisson bi-vector field directly on $\mathcal{M}$.
What we can do, is to induce a Poisson bi-vector field on the whole $\tilde{\mathcal{M}}$ and evaluate the associate bracket (which will obey the Leibnitz rule and the Jacobi identity) on the subalgebra of functions on $\tilde{\mathcal{M}}$ being pull-back via $\tau$ of functions on $\mathcal{M}$.
As discussed in section \ref{Subsubsec:The non-closed case} this bracket does not restrict to such a subalgebra and, in general, will present anomalies terms depending on the additional variables $\mu$.

First of all, the basis of differential one-forms on $\mathcal{M}$ dual to the basis of vector fields $\{Z_j,\, B_j\}_{j=1,2,3}$ is readily computed to be made by $\{\eta_j,\, \rho_j\}_{j=1,2,3}$ where $\eta_j$ are the right-invariant differential forms and $\rho_j \,=\, \dd \beta_j - \frac{1}{2} \epsilon_{jkl} \eta_k \wedge \eta_l$.
Using the basis $\{\eta_j,\, \rho_j,\, \theta_3,\, \dd \mu\}_{j=1,2,3}$ of differential one-forms on $\tilde{\mathcal{M}}$, the symplectic structure $\Omega$ reads
\be
\Omega \,=\, \delta^{jk} \rho_j \wedge \eta_k - \frac{1}{2} \epsilon_{j}^{\ kl} \left( \beta_l + (n-\mu)\theta_3^l \right) \eta_j \wedge \eta_k + \dd \mu \wedge \theta_3
\ee
where $\theta_3^l \,=\, \theta_3(\bar{Y}_l^{\uparrow})$.
The inverse of such a differential two-form is computed to be the following Poisson bi-vector field
\be
\Lambda \,=\, \delta^{jk} Z_j \wedge B_k - \frac{1}{2} \epsilon_{jkl} \left( \beta_l + (n - \mu) \theta_3^l \right) B_j \wedge B_k + \frac{\de}{\de \mu} \wedge \bar{X}_3^\uparrow \,.
\ee
That this Poisson bracket do not restrict to the subalgebra of functions on $\tilde{\mathcal{M}}$ being the pull-back of functions on $\mathcal{M}$ is readily seen by noting that the term $B_j \wedge B_k$ explicitly depends on $\mu$ and, thus, $\Lambda$ is not projectable to $\mathcal{M}$. 

As an example, we explicitly write the Poisson algebra closed by the pull-back via $\tau$ of following functions
\be
J_k \,=\, \beta_k - n \theta_3^k \,,
\ee
which reads
\be
\{\tilde{J}_j,\, \tilde{J}_k\} \,=\, \epsilon_{jkl}\tilde{J}_l + \mu \epsilon_{jkl} \theta_3^l \,.
\ee
where $\tilde{J}_k \,=\, \tau^\star J_k$.
\begin{remark}
The functions $J_k$ above are the restriction to the zero section of $\tau$ of the conserved currents associated with the canonical lift of the left-action of $SU(2)$ on $\mathcal{M}$. 
\end{remark}

\subsection{A second example: free Electrodynamics}
\label{Subsec:A second example: free Electrodynamics}

As extensively described in Ex. $2.7$ and Sect. $5.3$ of \cite{Ciaglia-DC-Ibort-Marmo-Schiav-Zamp2021-Cov_brackets_toappear}, a multisymplectic formulation of free Electrodynamics on the Minkowski space-time leads to a final stable manifold being:
\be
\mathcal{M}_\infty \,=\, \prod_{k=1,2,3} \mathcal{H}^{\frac{3}{2}}(\Sigma,\, vol_\Sigma)_k \times \prod_{k=1,2,3} {\mathcal{H}^{\frac{1}{2}}(\mathrm{div}0; \, \Sigma,\, vol_\Sigma)}^k \,,
\ee
whose points are denoted, in a system of local coordinates, by $m_\infty \,=\, (a_k,\, p^k)$ and where $\Sigma$ is any $1$-codimension hypersurface of the Minkowski space-time.

$\mathcal{M}_\infty$ is a pre-symplectic manifold equipped with the pre-symplectic structure:
\be
\os_\infty (\mathbb{X},\, \mathbb{Y}) \,=\, \int_\Sigma \left(\, {\mathbb{X}_a}_k {\mathbb{Y}_p}^k - {\mathbb{X}_p}^k {\mathbb{Y}_a}_k \,\right) vol_\Sigma \,,
\ee
whose kernel, at each point $m_\infty$ reads:
\be
K_{m_\infty} \,=\, < \left\{\, \mathbb{X} \;\; :\;\; {\mathbb{X}_a}_k \,=\, \de_k \phi \,\right\} > \,,
\ee
for some $\psi \in \mathcal{H}^{\frac{5}{2}}(\Sigma,\, vol_\Sigma)$.

Denoting by $K$ the characteristic distribution of $\os_\infty$, on the bundle $\mathcal{M}_\infty \to \mathcal{M}_\infty / K$ the connection represented by the following $1-1$ tensor field can be fixed:
\be
\mathcal{P} \,=\, P_k \otimes \frac{\delta}{\delta a_k} \,,
\ee
where by $\left\{\, \frac{\delta}{\delta a_k},\, \frac{\delta}{\delta p^k} \,\right\}_{k=1,2,3}$ we denote a basis of vector field over $\mathcal{M}_\infty$ and where:
\be
P_k \,=\, \de_k \int_\Sigma G_\Delta(\underline{x},\, \underline{y}) \delta^{jl} \de_j \delta a_l(y) \, \dd^3 y   \,,
\ee
with $G_\Delta$ being the Green's function of the Laplacian operator, $\underline{x}$, $\underline{y}$ points in $\Sigma$ and $\{\, \delta a_j \,\}_{j=1,2,3}$ a dual basis of $\left\{\, \frac{\delta}{\delta a_j} \,\right\}_{j=1,2,3}$.
Such a connection, gives rise to the following splitting of $\mathbf{T}_{m_\infty} \mathcal{M}_\infty \simeq \mathcal{M}_\infty$:
\be
\mathbf{T}_{m_\infty} \mathcal{M}_\infty \simeq \mathcal{M}_\infty \,=\, \prod_{k=1,2,3} {\mathcal{H}^{\frac{3}{2}}(\mathrm{div}0; \,\Sigma,\, vol_\Sigma)}_k \oplus \mathrm{grad} \mathcal{H}^{\frac{5}{2}} \times \prod_{k=1,2,3} {\mathcal{H}^{\frac{1}{2}}(\mathrm{div}0;\, \Sigma,\, vol_\Sigma)}^k \,, 
\ee
where the complement of $K_{m_\infty} \,=\, \mathrm{grad} \mathcal{H}^{\frac{5}{2}}$ represents the horizontal distribution associated with the connection.

The enlarged manifold obtained extending $\mathcal{M}_\infty$ via the dual of $K_{m_\infty}$ in the spirit of the coisotropic embedding theorem is:
\be 
\begin{split}
\tilde{\mathcal{M}} \,&\simeq \mathbf{T}_{\tilde{m}} \tilde{\mathcal{M}}\,=\, \prod_{k=1,2,3} {\mathcal{H}^{\frac{3}{2}}(\mathrm{div}0; \,\Sigma,\, vol_\Sigma)}_k \oplus \mathrm{grad} \mathcal{H}^{\frac{5}{2}} \times \prod_{k=1,2,3} {\mathcal{H}^{\frac{1}{2}}(\mathrm{div}0;\, \Sigma,\, vol_\Sigma)}^k \times \mathrm{grad} {\mathcal{H}^{\frac{5}{2}}}^\star \, \simeq \\
&\simeq \, \prod_{k=1,2,3} {\mathcal{H}^{\frac{3}{2}}(\mathrm{div}0; \,\Sigma,\, vol_\Sigma)}_k \oplus \mathrm{grad} \mathcal{H}^{\frac{5}{2}} \times \prod_{k=1,2,3} {\mathcal{H}^{\frac{1}{2}}(\mathrm{div}0;\, \Sigma,\, vol_\Sigma)}^k \times \mathrm{grad} \mathcal{H}^{\frac{5}{2}} \,,
\end{split}
\ee
where a point will be denoted, in a system of local coordinates, by $\tilde{m} \,=\, (\tilde{a}_k,\, \de_k \psi,\, p^k,\, \mu^k)$.

In this case, a direct computation shows that:
\be
\dd P_k \,=\, 0 
\ee
and, thus, we are in the case of Sect. \ref{Subsubsec:The closed case}.
Consequently, denoting an element of $\mathbf{T}_{\tilde{m}} \tilde{\mathcal{M}}$ by $\mathbb{X} \,=\, ({\tilde{\mathbb{X}}_a}_k,\, \de_k \mathbb{X}_\psi,\, {\mathbb{X}_p}^k,\, {\mathbb{X}_\mu}^k)$ the symplectic structure on $\tilde{\mathcal{M}}_\infty$ is:
\be
\tilde{\Omega}(\mathbb{X},\, \mathbb{Y}) \,=\, \underbrace{\int_\Sigma \left(\, {\tilde{\mathbb{X}}_a}_k {\mathbb{Y}_p}^k - {\mathbb{X}_p}^k {\tilde{\mathbb{X}}_a}_k \,\right) vol_\Sigma}_{= \Omega_W(\mathbb{X},\, \mathbb{Y})} + \underbrace{{\mathbb{X}_\mu}^k \de_k \mathbb{Y}_\psi - \de_k \mathbb{X}_\psi {\mathbb{Y}_\mu}^k}_{= \Omega_{K \oplus K^\star}(\mathbb{X},\, \mathbb{Y})} \,.
\ee
The inverse of $\tilde{\Omega}$ reads:
\be
\tilde{\Lambda} \,=\, \Lambda_W \oplus \Lambda_{K \oplus K^\star} \,,
\ee
where:
\be
\Lambda_W \,=\, \frac{\delta}{\delta \tilde{a}_k} \wedge \frac{\delta}{\delta p^k} \,,
\ee
with an integration over $\Sigma$ implied in the wedge product.
Such a bivector field is a Poisson bivector field being the inverse of a closed and non-degenerate form on $W$ and it is projectable to the Poisson bivector field over $\mathcal{M}_\infty$:
\be
\lambda \,=\, \frac{\delta}{\delta \tilde{a}_k} \wedge \frac{\delta}{\delta p^k} \,,
\ee
giving rise to the following Poisson bracket between any two functions on $\mathcal{M}_\infty$:
\be
\left\{\,f,\, g\,\right\} \,=\, \int_\Sigma \left(\, \frac{\delta f}{\delta \tilde{a}_k} \frac{\delta g}{\delta p^k} - \frac{\delta f}{\delta p^k} \frac{\delta g}{\delta \tilde{a}_k} \,\right) vol_\Sigma \,,
\ee
which coincides with the bracket in Eq. $145$ of \cite{Ciaglia-DC-Ibort-Marmo-Schiav-Zamp2021-Cov_brackets_toappear}.

\subsection{Multisymplectic formulation of Yang-Mills theories}
\label{Subsec:Multisymplectic formulation of Yang-Mills theories}

Let us now pass to the main aim of the paper, that is the construction of the Poisson structure in the case of Yang-Mills theories. 

Let us start by describing the multisymplectic formulation of Yang-Mills theories on the Minkowski space-time.
Here, the space-time reads $\m \,=\, (\mathbb{R}^4,\, \eta)$, $\eta$ being the Minkowski metric.
Yang-Mills fields are connection one-forms on a principal fibre bundle with structure group $G$ whose Lie algebra will be denoted by $\mathfrak{g}$.
Usually $G$ is a compact and connected group, indeed, the most relevant examples of Yang-Mills type theories are Electrodynamics, the theory of weak interactions and Quantum Chromodynamics of strong interactions where $G$ is respectively $U(1)$, $SU(2)$ and $SU(3)$.
Their unification is, indeed, described as a Yang-Mills theory with $G\,=\, U(1) \times SU(2) \times SU(3)$.
Connection one-forms are differential one forms on $\m$ with values in $\mathfrak{g}$.
Thus, the fibre bundle $\mathbb{E}$ underlying the theory is $\pi \;:\; \mathbb{E} \,=\, \mathbf{T}^\star \m \otimes \mathfrak{g} \to \m$ and Yang-Mills fields are sections of $\pi$.
We denote by $\left(\, x^\mu,\, \alpha^a_\mu \,\right) ,\, \mu=0,...,3,\, a=1,..., dim\mathfrak{g}$ a system of local coordinates on $\mathbb{E}$ defined on an open set $U_\mathbb{E} \subset \mathbb{E}$, and we denote sections of $\pi$ by:
\be
A \;\; :\;\; U_\m \to U_\mathbb{E} \;\; :\; x^\mu \mapsto A_\mu^a(x) \dd x^\mu \otimes \xi_a \,,
\ee
where $U_\m$ is an open set in $\m$ and $\{\, \xi_a \,\}_{a=1,...,dim\mathfrak{g}}$ is a basis of $\mathfrak{g}$.

The Covariant Phase Space $\pe$ in this case is $\mathbf{T}\m \otimes \mathbf{T}\m \otimes \mathfrak{g}^\star \otimes \mathbf{T}^\star \m \otimes \mathfrak{g}$.
However, in order to obtain the correct equations of the motion, we will restrict\footnote{This restriction is allowed because the manifold we will obtain, even if it is not strictly speaking the reduced dual of $\pi$, is still a multisymplectic manifold.} to the subbundle $\mathbf{T}\m \wedge \mathbf{T}\m \otimes \mathfrak{g}^\star \otimes \mathbf{T}^\star \m \otimes \mathfrak{g}$ whose sections are couples of connection one-forms and bivector fields with values in $\mathfrak{g}^\star$.
We denote by $\left(\, x^\mu,\, \alpha_\mu^a, \, \rho^{\mu \nu}_a \, \right),\, \mu,\, \nu=0,...,3,\, a=1,...,dim\mathfrak{g}$ a system of local coordinates on $\pe$ defined on an open set $U_{\pe} \subset
\pe$.
Sections of $\delta_1 \; :\; \pe \to \m$ will be denoted by:
\be
\chi \,=\, (A,\, P) \;\; :\;\; U_\m \to U_{\pe} \;\; :\;\; x^\mu \mapsto \left(\, A_\mu^a(x)\dd x^\mu \otimes \xi_a ,\, P^{\mu \nu}_a (x, A(x)) \frac{\de}{\de x^\mu} \wedge \frac{\de }{\de x^\nu} \otimes \xi^a \,\right) \,,
\ee
where $U_{\m}$ is an open set in $\m$ and $\{\, \xi^a \, \}_{a=1,...,dim\mathfrak{g}}$ is a basis of $\mathfrak{g}^\star$.

The Yang-Mills Hamiltonian $H$ is:
\be \label{Eq:yang-mills hamiltonian}
H(x, \alpha,\, \rho) \,=\, \frac{1}{4}\rho^{\mu\nu}_a \rho_{\mu \nu}^a + \frac{1}{2} \epsilon^a_{bc} \rho^{\mu \nu}_a \alpha^b_\mu \alpha^c_\nu \,,
\ee
where space-time indices are raised/lowered by means of the Minkowski metric and Lie algebra indices by means of the Killing-Cartan metric on $\mathfrak{g}$ and where $\epsilon^a_{bc}$ are the structure constants of the Lie algebra $\mathfrak{g}$.
With this Hamiltonian, the action functional is seen to be:
\be \label{Eq: yang-mills action}
\ac_\chi \,=\, \int_\m \chi^\star \Theta_H \,=\, -\int_\m \left[\, P^{\mu \nu}_a F^a_{\mu \nu} + \frac{1}{4} P^{\mu \nu}_a P_{\mu \nu}^a  \,\right] vol_\m \,,
\ee
where: 
\be
F_{\mu \nu}^a \,=\, \frac{1}{2} \left(\, \de_\mu A_\nu^a - \de_\nu A_\mu^a + \epsilon^a_{bc} A^b_\mu A^c_\nu \,\right) \,=\, (\nabla A)^a_{\mu \nu} \,=:\, \nabla_\mu A_\nu^a \,,
\ee
are the coefficients of the curvature of the connection $A$ and where $\nabla$ represents the covariant derivative with respect to the connection $A$.

For technical reasons that will be clear in the sequel, we will chose the space of dynamical fields made by $\mathcal{H}^3$ potentials and $\mathcal{H}^1$ momenta fields.
The construction of the space of dynamical fields goes as follows.
We consider the space $\Gamma_{\textsc{split}}^0(\delta_1)$ of smooth, splitting sections of $\delta_1$ with compact support, on which the action functional \ref{Eq: yang-mills action} is well defined.
On $\Gamma_{\textsc{split}}^0(\delta_1)$ we consider the norm $\Vert \chi \Vert_{\mathscr{C}} \,=\, \sum_{\mu, a} \Vert A_\mu^a \Vert_{\mathcal{H}^3} + \sum_{\mu, \nu, a}\Vert P^{\mu \nu}_a \Vert_{\mathcal{H}^1}$.
As the following proposition proves, $\ac$ is continuous in the norm $\Vert \, \cdot \, \Vert_{\mathscr{C}}$ and, therefore, it can be extended by continuity to the closure of $\Gamma_{\textsc{split}}^0(\delta_1)$ with respect to $\Vert \, \cdot \, \Vert_{\mathscr{C}}$, i. e., to:
\be
\overline{\Gamma_{\textsc{split}}^0(\delta_1)}^{\Vert \, \cdot \, \Vert_{\mathscr{C}}} \,=:\, \fpe \,=\, \prod_{\mu, a} \mathcal{H}^3(\m, \, vol_\m)^a_\mu \times \prod_{\mu, \nu, a} \mathcal{H}^2(\m, \, vol_\m)^{\mu \nu}_a \,.
\ee
Now, let us prove the continuity of $\ac$.
\begin{proposition}[\textsc{Continuity of $\ac$}]
The action functional:
\be
\ac_\chi \,=\, -\int_\m \left[\, P^{\mu \nu}_a F^a_{\mu \nu} + \frac{1}{4} P^{\mu \nu}_a P_{\mu \nu}^a  \,\right] vol_\m \,,
\ee
is well defined on $\Gamma^0_{\textsc{split}}(\delta_1)$ and is continuous in the norm $\Vert \, \cdot \, \Vert_{\mathscr{C}}$.
\begin{proof}
That $\ac$ is well defined on $\Gamma^0_{\textsc{split}}(\delta_1)$ is obvious.

Regarding the continuity, the following estimate holds:
\be
\begin{split}
\vert \, \ac_\chi - \ac_{\tilde{\chi}} \, \vert  \,&=\, \left\vert\, \int_{\m} \left( P^{\mu \nu}_a F^a_{\mu \nu} + \frac{1}{4} P^{\mu \nu}_a P_{\mu \nu}^a - \tilde{P}^{\mu \nu}_a \tilde{F}^a_{\mu \nu} - \frac{1}{4} \tilde{P}^{\mu \nu}_a \tilde{P}^a_{\mu \nu} \right) vol_{\m} \, \right\vert  \, \leq \, \\
\, &\leq \, \underbrace{\int_\m \left\vert (P^{\mu \nu}_a - \tilde{P}^{\mu \nu}_a) F^a_{\mu \nu}   \right\vert vol_\m}_{\mathscr{I}_1} + \underbrace{\int_\m \left\vert \tilde{P}^{\mu \nu}_a (F^a_{\mu \nu} - \tilde{F}^a_{\mu \nu}) \right\vert vol_\m}_{\mathscr{I}_2} + \\
&+ \, \underbrace{\frac{1}{4} \int_\m \left\vert P^{\mu \nu}_a (P^a_{\mu \nu} - \tilde{P}^a_{\mu \nu}) \right\vert vol_\m}_{\mathscr{I}_3} + \underbrace{\frac{1}{4}\int_\m \left\vert \tilde{P}^{\mu \nu}_a (P^a_{\mu \nu} - \tilde{P}^a_{\mu \nu}) \right\vert vol_\m}_{\mathscr{I}_4} \,,
\end{split}
\ee
where $\tilde{\chi} \,=\, (\tilde{A}, \tilde{P})$ and $\tilde{F}^a_{\mu \nu} \,=\, \frac{1}{2} \left(\de_\mu \tilde{A}_\nu^a - \de_\nu \tilde{A}_\mu^a + \epsilon^a_{\ bc} \tilde{A}_\mu^b \tilde{A}_\nu^c  \right)$.
Now:
\be
\mathscr{I}_1 \,\leq\, \sum_{\mu, \nu, a} \Vert P^{\mu \nu}_a - \tilde{P}^{\mu \nu}_a \Vert_{\mathcal{L}^2} \, \Vert F^a_{\mu \nu} \Vert_{\mathcal{L}^{2}} \,\leq \, \sum_{\mu, \nu, a} \Vert P^{\mu \nu}_a - \tilde{P}^{\mu \nu}_a \Vert_{\mathcal{H}^2} \, \Vert F^a_{\mu \nu} \Vert_{\mathcal{L}^{2}} \,,
\ee
\be
\begin{split}
\mathscr{I}_2 \,&\leq\, \sum_{\mu, \nu, a} \Vert \tilde{P}^{\mu \nu}_a \Vert_{\mathcal{L}^2} \, \frac{1}{2} \bigl( \Vert \de_\mu (A_\nu^a - \tilde{A}_\nu^a) \Vert_{\mathcal{L}^2} +  \Vert \de_\nu (A_\mu^a - \tilde{A}_\mu^a) \Vert_{\mathcal{L}^2} \bigr) + \\ 
&+ \sum_{\mu, \nu, a} \frac{1}{2} \left( \vert \epsilon^a_{\ bc} \vert \Vert \tilde{P}^{\mu \nu}_a A_\mu^b \Vert_{\mathcal{L}^2} \Vert A_\nu^c - \tilde{A}_\nu^c \Vert_{\mathcal{L}^2} + \vert \epsilon^a_{\ bc} \vert \Vert \tilde{P}^{\mu \nu}_a A_\nu^c \Vert_{\mathcal{L}^2} \Vert A_\mu^b - \tilde{A}_\mu^b \Vert_{\mathcal{L}^2} \right) \,\leq \, \\
\,&\leq\, \sum_{\mu, \nu, a} \left( \Vert \tilde{P}^{\mu \nu}_a \Vert_{\mathcal{L}^2} + \vert \epsilon^a_{\ bc} \vert \Vert \tilde{P}^{\mu \nu}_a A_\mu^b \Vert_{\mathcal{L}^2}  \right) \Vert A_\nu^c - \tilde{A}_\nu^c \Vert_{\mathcal{H}^3}   
\end{split}
\ee
\be
\mathscr{I}_3 \,\leq \, \frac{1}{4} \sum_{\mu, \nu, a} \Vert P^{\mu \nu}_a \Vert_{\mathcal{L}^2} \Vert P_{\mu \nu}^a - \tilde{P}_{\mu \nu}^a \Vert_{\mathcal{L}^2} \, \leq \, \frac{1}{4} \sum_{\mu, \nu, a} \Vert P^{\mu \nu}_a \Vert_{\mathcal{L}^2} \Vert P_{\mu \nu}^a - \tilde{P}_{\mu \nu}^a \Vert_{\mathcal{H}^2}
\ee
\be
\mathscr{I}_4 \,\leq\, \frac{1}{4} \sum_{\mu, \nu, a} \Vert \tilde{P}^{\mu \nu}_a \Vert_{\mathcal{L}^2} \Vert P_{\mu \nu}^a - \tilde{P}_{\mu \nu}^a \Vert_{\mathcal{L}^2} \, \leq \, \frac{1}{4} \sum_{\mu, \nu, a} \Vert \tilde{P}^{\mu \nu}_a \Vert_{\mathcal{L}^2} \Vert P_{\mu \nu}^a - \tilde{P}_{\mu \nu}^a \Vert_{\mathcal{H}^2} \,.
\ee
Therefore, it is evident that if $\Vert \chi - \tilde{\chi} \Vert_{\mathscr{C}} \to 0$, then, since both $\Vert A_\mu^a - \tilde{A}_\mu^a \Vert_{\mathcal{H}^3}$ and $\Vert P^{\mu \nu}_a - \tilde{P}^{\mu \nu}_a \Vert_{\mathcal{H}^2}$ goes to zero for all $\mu, \nu, a$, $\left\vert \ac_\chi - \ac_{\tilde{\chi}} \right\vert \to 0$.
\end{proof}
\end{proposition}

\begin{remark}
Actually, the regularity chosen is not the minimal requirement for the $A_\mu$'s, since we really only need their covariant derivatives to be square integrable whereas the functions themselves may be not integrable still leaving the action functional well defined.
We restrict to this class of functions, since it is very useful from the mathematical point of view.
However, physically speaking, it must be clear that this exclude some (actually, never observed in nature) situations, such as the magnetic monopole, where the potential $A^a_\mu$ is non-zero at infinity.  
\end{remark}
Thus, our space of dynamical fields reads:
\be
\fpe \,=\, \prod_{a, \mu} {\mathcal{H}^3(\m,\, vol_\m )}^a_\mu \times \prod_{a,\, \mu,\,\nu} {\mathcal{H}^2(\m,\, vol_\m)}^{\mu \nu}_a \,,
\ee
which is a Hilbert manifold (actually an Hilbert space) whose tangent space at each point reads:
\be
\mathbf{T}_\chi \fpe \,=\, \prod_{a, \mu} {\mathcal{H}^3(\m,\, vol_\m )}^a_\mu \times \prod_{a,\, \mu,\,\nu} {\mathcal{H}^2(\m,\, vol_\m)}^{\mu \nu}_a \,.
\ee

Now, let us consider the slice $\Sigma \,=\, \left\{\, m \in \m \;\; :\;\; x^0 \,=\, x^0_\Sigma \,\right\}$.
By virtue of the trace theorem \cite{Lions1990-vol2} the space $\fpe^\Sigma$ turns out to be:
\be
\fpe^\Sigma \,=\, \prod_{a, \mu} {\mathcal{H}^{\frac{5}{2}}(\Sigma,\, vol_\Sigma)}^a_\mu \times \prod_{a, \mu, \nu} {\mathcal{H}^{\frac{3}{2}}(\Sigma,\, vol_\Sigma)}^{\mu \nu}_a \,. 
\ee
Elements is $\fpe^\Sigma$, will be denoted by $\chi_\Sigma$ and we will use the following coordinate expression for them:
\be
\chi_\Sigma \,=\, \left(\, A_\mu^a\bigr|_\Sigma(\underline{x}),\, P^{k0}_a\bigr|_\Sigma (\underline{x}),\, P^{jk}_a\bigr|_\Sigma (\underline{x}) \,\right) \,=:\, \left(\,a_k^a(\underline{x}),\, p^k_a(\underline{x}),\, \beta^{jk}_a(\underline{x})\,\right) \,,
\ee
where $\underline{x}$ is the coordinatization of a point in $\Sigma$.
Again, being $\fpe^\Sigma$ a Hilbert manifold and, actually a Hilbert space, its tangent space at each point is isomorphic to the Hilbert space itself:
\be
\mathbf{T}_{\chi_\Sigma} \fpe^\Sigma \,=\, \prod_{a, \mu} {\mathcal{H}^{\frac{5}{2}}(\Sigma,\, vol_\Sigma)}^a_\mu \times \prod_{a, \mu, \nu} {\mathcal{H}^{\frac{3}{2}}(\Sigma,\, vol_\Sigma)}^{\mu \nu}_a \,. 
\ee

Now, the $2$-form $\os$ reads:
\be \label{Eq:os Yang-Mills}
\os(\mathbb{X}_{\chi_\Sigma},\, \mathbb{Y}_{\chi_\Sigma}) \,=\, \int_\Sigma \left[\, {\mathbb{X}_a}^a_k {\mathbb{Y}_p}^k_a - {\mathbb{X}_p}_a^k {\mathbb{Y}_a}_k^a \,\right] vol_\Sigma \,,
\ee
where ${\mathbb{X}_a}^a_k$ and ${\mathbb{X}_p}_a^k$ are the $a_k^a$ and the $p_a^k$ components of a tangent vector to $\fpe^\Sigma$ at $\chi_\Sigma$.
On the other hand, the Hamiltonian functional is:
\be
\mathcal{H}(\chi_\Sigma) \,=\, \int_\Sigma \left[\, \frac{1}{2} p^k_a p^a_k + \frac{1}{4} \beta^{jk}_a \beta^a_{jk} + p^k_a \nabla_k a_0^a + \frac{1}{2} \beta^{jk}_a \nabla_j a_k^a   \,\right] vol_\Sigma \,,
\ee
where:
\be
\begin{split}
\nabla_k a_0^a \,&=\, \de_k a_0 + \epsilon^a_{bc} a_0^b a_k^c \,, \\
\nabla_j a_k^a \,&=\, \frac{1}{2}(\de_j a_k^a - \de_k a_j^a + \epsilon^a_{bc}a^b_j a^c_k ) \,.
\end{split}
\ee

Now, we use the PCA to find solutions of the pre-symplectic system $(\fpe^\Sigma,\, \os,\, \mathcal{H})$.
Following what we recalled in Appendix $A1$ of \cite{Ciaglia-DC-Ibort-Marmo-Schiav-Zamp2021-Cov_brackets_toappear}, first we should determine ${\mathbf{T}_{\chi_\Sigma} \fpe^{\Sigma}}^\perp \,=\, \mathrm{ker} \os$.
By looking at Eq. \eqref{Eq:os Yang-Mills}, $\mathrm{ker}\os$ is made by tangent vectors having only components along $a_0^a$ and by tangent vectors having only components along $\beta^{jk}_a$.
Let us denote a basis of $\mathrm{ker} \os$ by:
\be
\left\{\, \frac{\delta }{\delta a_0^a},\, \frac{\delta}{\delta \beta^{jk}_a} \,\right\}_{a=1,...,dim\mathfrak{g},\, j,k=1,...,3} \,.
\ee
Then, the first manifold of the PCA is obtained by imposing that such tangent vectors lie also in the kernel of $\dd \mathcal{H}$ at each $\chi_\Sigma$.
The following conditions emerge:
\be \label{Eq:first step pca Yang-Mills}
\begin{split}
i_{\frac{\delta}{\delta a_0^a}} \dd \mathcal{H} \,=\, 0 \;\; &\implies \;\; \nabla^\star_k p^k_a \,=\, 0 \,, \\
i_{\frac{\delta}{\delta \beta^{jk}_a}}\dd \mathcal{H} \,=\, 0 \;\; &\implies \;\; \beta^{jk}_a \,=\, - \eta^{jl}\eta^{km} \nabla_l a_m^a \,,
\end{split}
\ee
where $\nabla^\star_k p^k_a \,=\, \de_k p^k_a + {{\epsilon_a}^b}_c p_b^k a^c_k $ is the adjoint of the covariant differential.
The second constraint can be eliminated by replacing $\beta^{jk}_a$ with the expression on the right hand side in terms of the $a_k^a$ while the first constraint can not be eliminated.
Therefore, the first manifold obtained via the PCA is:
\be
\mathcal{M}_1 \,=\, \left\{\, (a_k^a,\, p^k_a) \;\; :\;\; \nabla^\star_k p^k_a \,=\, 0 \,\right\} \,=:\, \prod_{k, a} {\mathcal{H}^{\frac{5}{2}}(\Sigma,\, vol_\Sigma)}^a_k \times \prod_{k, a} {\mathcal{H}^{\frac{3}{2}}(\Sigma,\, \nabla_k^\star p^k_a = 0,\, vol_\Sigma)}^k_a \,,
\ee
which can be immersed into $\fpe^\Sigma$ via the second of Eq. \eqref{Eq:first step pca Yang-Mills} and by fixing any arbitrary $a_0^a$.
We denote such immersion by $\mathfrak{i}_1$.
We will prove that $\mathcal{M}_1$ is actually a Hilbert space itself by proving that it is a closed subspace of $\fpe^{\Sigma}$.
But, for the moment let us proceed with the PCA.
In the second step of the PCA we should determine ${\mathbf{T}_{\chi_\Sigma} \mathfrak{i}_1 (\mathcal{M}_1)}^\perp$.
It is made by tangent vectors $\mathbb{X}_{\chi_\Sigma}$ to $\fpe^{\Sigma}$ at $\chi_\Sigma \in \mathfrak{i}_1(\mathcal{M}_1)$ such that $\mathfrak{i}^\star_1 (i_{\mathbb{X}_{\chi_\Sigma}} \os) \,=\,0$.
It is readily seen that by virtue of the constraint $\nabla^\star_k p^k_a \,=\,0$, ${\mathbf{T}_{\chi_\Sigma} \mathfrak{i}_1 (\mathcal{M}_1)}^\perp$ is made by tangent vectors having $a_k^a$ component equal to the covariant differential of a Lie algebra-valued function, say $\psi^a$\footnote{We will chose $\psi^a$ to lie in $\mathcal{H}^{\frac{7}{2}}(\Sigma,\, vol_\Sigma)$ so that we are ensured that its covariant gradient is again a $\mathcal{H}^{\frac{5}{2}}(\Sigma,\, vol_\Sigma)$ function.
Indeed, this is due to the fact that $\mathcal{H}^{\frac{5}{2}}(\Sigma, vol_\Sigma)$ is a Banach algebra (see \cite{Adams1975}) which ensures the following inequality
\be
\sum_{k,a} \Vert \nabla_k \psi^a \Vert_{\mathcal{H}^{\frac{5}{2}}} \,\leq \, \sum_{k, a} \Vert \de_k \psi^a \Vert_{\mathcal{H}^{\frac{5}{2}}} + \sum_{k, a} \vert \epsilon^a_{\ bc} \vert \Vert a_k^b \psi^c \Vert_{\frac{5}{2}} \, \leq \, 3 \sum_{k, a} \Vert \psi^a \Vert_{\mathcal{H}^{\frac{7}{2}}} + \sum_{k, a} \vert \epsilon^a_{\ bc} \vert B_{b, c, k} \Vert a_k^b \Vert_{\mathcal{H}^{\frac{5}{2}}} \Vert \psi^c \Vert_{\frac{5}{2}} \,,
\ee
for some constants $B_{b,c,k}$, where the inequality $\sum_{k, a} \Vert \de_k \psi^a \Vert_{\mathcal{H}^{\frac{5}{2}}} \leq 3 \sum_{k, a} \Vert \psi^a \Vert_{\mathcal{H}^{\frac{7}{2}}}$ is due to
\be
\sum_{k, a} \Vert \de_k \psi^a \Vert_{\mathcal{H}^{\frac{5}{2}}} \,=\, \sum_{k,a} \left\Vert \vert k \vert^{\frac{5}{2}} k_k  \tilde{\psi}^a \right\Vert_{\mathcal{L}^2} \, \leq \, 3 \sum_{a} \left\Vert \vert k \vert^{\frac{7}{2}} \tilde{\psi}^a \right\Vert_{\mathcal{L}^2} \,=\, 3 \sum_a \Vert \psi^a \Vert_{\mathcal{H}^{\frac{7}{2}}} \,,
\ee
where $\tilde{\psi}^a$ is the Fourier transform of $\psi^a$.
}, and $p^k_a$ component equal to $[p^k, \psi]^a$, $[\,\cdot\,,\,\cdot\,]$ being the Lie bracket in $\mathfrak{g}$:
\be 
{\mathbf{T}_{\chi_\Sigma} \mathfrak{i}_1 (\mathcal{M}_1)}^\perp \,=\, \left\{\, \mathbb{X}_{\chi_\Sigma} \in \mathbf{T}_{\chi_\Sigma}\fpe^\Sigma\bigr|_{\mathfrak{i}^1(\mathcal{M}_1)} \;\; :\;\; {\mathbb{X}_a}_k^a \,=\, \nabla_k \psi^a \,, \; \; {\mathbb{X}_p}_a^k \,=\, [p^k,\, \psi]_a \,\right\} \,.
\ee
Note that, consistently, $\mathbb{X}_{\chi_\Sigma}$ is a tangent vector to $\mathfrak{i}_1(\mathcal{M}_1) \subset \fpe^\Sigma$ since $\nabla_k \psi^a$ is an ${\mathcal{H}^{\frac{5}{2}}(\Sigma,\, vol_\Sigma)}$ function and $[p^k,\, \psi]_a$ is an $\mathcal{H}^{\frac{7}{2}}(\Sigma,\, vol_\Sigma)$ function because $p^k_a \in {\mathcal{H}^{\frac{3}{2}}(\Sigma,\, vol_\Sigma)}^k_a \supset {\mathcal{H}^{-\frac{7}{2}}(\Sigma,\, vol_\Sigma)}^k_a \,=\, {{\mathcal{H}^{\frac{7}{2}}(\Sigma,\, vol_\Sigma)}^k_a}^\star$, thus: 
\be \label{Eq:commutator scalar product}
\int_\Sigma [p^k, \psi]_a vol_\Sigma \,=\, \int_\Sigma \epsilon_{a \ c}^{\ b} p^k_b \psi^c vol_\Sigma \,=\, \epsilon_{a \ c}^{\ b} \langle p^k_b ,\, \psi^c \rangle_{\mathcal{H}^{\frac{7}{2}}} \,,
\ee
and, consequently: 
\be \label{Eq:norm commutator}
\Vert [p^k,\, \psi]_a \Vert_{\mathcal{H}^{\frac{7}{2}}} \,\leq\, |\epsilon^{\ b}_{a \ c}|\Vert p^k_b \Vert_{H^{-\frac{7}{2}}} \Vert \psi^c \Vert_{\mathcal{H}^{\frac{7}{2}}} \leq \infty \,.
\ee
Let us denote elements of ${\mathbf{T}_{\chi_\Sigma} \mathfrak{i}_1 (\mathcal{M}_1)}^\perp$ by $\mathbb{X}^{\textsc{gauge}}_\psi$.
A direct computation shows that $i_{\mathbb{X}^{\textsc{gauge}}_\psi} \dd \mathcal{H} \,=\, 0 \,\, \forall\,\, \psi^a$, that is, the PCA stops after finding the first manifold $\mathcal{M}_1 \,=:\, \mathcal{M}_\infty$.

Therefore, the final manifold obtained out of the PCA is the Hilbert manifold:
\be
\mathcal{M}_\infty \,=\, \left\{\, (a_k^a,\, p^k_a) \;\; :\;\; \nabla^\star_k p^k_a \,=\, 0 \,\right\} \,=:\, \prod_{k, a} {\mathcal{H}^\frac{5}{2}(\Sigma,\, vol_\Sigma)}^a_k \times \prod_{k, a} {\mathcal{H}^\frac{3}{2}(\Sigma,\, \nabla_k^\star p^k_a = 0,\, vol_\Sigma)}^k_a \,.
\ee
That this is a Hilbert manifold and, actually a Hilbert space, is due to the fact that ${\mathcal{H}^\frac{5}{2}(\Sigma,\, vol_\Sigma)}^a_k$ is a Hilbert space and that ${\mathcal{H}^\frac{3}{2}(\Sigma,\, \nabla_k^\star p^k_a = 0,\, vol_\Sigma)}^k_a$ is a closed (and, thus, Hilbert) subspace of ${\mathcal{H}^1(\Sigma, vol_\Sigma)}^k_a$.
Let us prove this last claim.
Being $\nabla$ the covariant derivative associated with a $\mathcal{H}^\frac{5}{2}$ connection, it acts as a linear operator between the Hilbert spaces ${\mathcal{H}^\frac{5}{2}(\Sigma,\, vol_\Sigma)}^a$ and ${\mathcal{H}^{\frac{3}{2}}(\Sigma,\, vol_\Sigma)}^a_k$:
\be
\nabla \;\; :\;\; {\mathcal{H}^\frac{5}{2}(\Sigma,\, vol_\Sigma)}^a \to {\mathcal{H}^{\frac{3}{2}}(\Sigma,\, vol_\Sigma)}^a_k \;\; :\;\; f^a \mapsto \nabla_k f^a \,.
\ee
Its adjoint is an operator from ${{\mathcal{H}^{\frac{3}{2}}(\Sigma,\, vol_\Sigma)}^k_a}^\star$ that can be identified with ${\mathcal{H}^\frac{3}{2}(\Sigma,\, vol_\Sigma)}^a_k$ itself, to ${{\mathcal{H}^\frac{5}{2}(\Sigma,\, vol_\Sigma)}^a}^\star$ that can be identified with ${\mathcal{H}^\frac{5}{2}(\Sigma,\, vol_\Sigma)}_a$:
\be
\nabla^\star \;\; :\;\; {\mathcal{H}^\frac{3}{2}(\Sigma,\, vol_\Sigma)}^k_a \to {\mathcal{H}^\frac{5}{2}(\Sigma,\, vol_\Sigma)}_a \;\; :\;\; p^k_a \mapsto \nabla^\star_k p^k_a \,.  
\ee
The following holds.
\begin{proposition}[\textsc{Closedness of $\nabla$ in $\mathcal{H}^\frac{3}{2}$}] \label{Prop: closedness nabla in H3/2}
The operator $\nabla$ is a closed operator from ${\mathcal{H}^{\frac{5}{2}}(\Sigma,\, vol_\Sigma)}^a$ to ${\mathcal{H}^{\frac{3}{2}}(\Sigma,\, vol_\Sigma)}_k^a$.
\begin{proof}
Consider a sequence of functions in ${\mathcal{H}^\frac{5}{2}(\Sigma, vol_\Sigma)}^a$, say $\left\{\, f^a_n \,\right\}_{n \in \mathbb{N}}$ converging to some $f^a \in {\mathcal{H}^\frac{5}{2}(\Sigma, vol_\Sigma)}^a$ in the $\mathcal{H}^\frac{5}{2}$-norm.
Then $\nabla_k f^a_n$ converges to $\nabla_k f^a$ in the $\mathcal{H}^{\frac{3}{2}}$-norm. 
Indeed:
\be
\begin{split}
\sum_{k, a} \Vert \nabla_k f^a_n - \nabla_k f^a \Vert_{\mathcal{H}^{\frac{3}{2}}} \,&=\, \sum_{k, a} \Vert \nabla_k (f^a_n - f^a) \Vert_{\mathcal{H}^{\frac{3}{2}}}  \,=\,  \sum_{k, a}\Vert \de_k (f_n^a - f^a) + \epsilon^a_{\ bc} a^b_k (f^c_n - f^c) \Vert_{\mathcal{H}^{\frac{3}{2}}} \,=\, \\
\,&\leq \, \sum_{k,a } \Vert \de_k (f_n^a - f^a) \Vert_{\mathcal{H}^{\frac{3}{2}}} +  \sum_{k, a} + \sum_{k,a} \vert \epsilon^a_{\ bc} \vert \Vert a_k^b (f_n^c - f^c) \Vert_{\mathcal{H}^{\frac{3}{2}}} \, \leq \\ 
\, &\leq \, 3 \sum_{a} \Vert f^a_n - f^a \Vert_{\mathcal{H}^{\frac{5}{2}}} + \sum_{k,a} \vert \epsilon^a_{\ bc} \vert \Vert a_k^b \Vert_{\mathcal{H}^{\frac{5}{2}}} \, \Vert f^c_n - f^c \Vert_{\mathcal{H}^{\frac{5}{2}}} \,,
\end{split}
\ee
where the last inequality is due to the content of footnote $13$ and to the fact that $\mathcal{H}^{\frac{5}{2}}(\Sigma, vol_\Sigma)$ is a Banach algebra.
Because of the latter inequality, $\sum_{k, a} \Vert \nabla_k f^a_n - \nabla_k f^a \Vert_{\mathcal{H}^{\frac{3}{3}}}$ approaches zero when $\Vert f^c_n - f^c \Vert_{\mathcal{H}^{\frac{5}{2}}}$ approaches zero.
Thus, by definition of closed operator, $\nabla$ is closed.
\end{proof}
\end{proposition}
Therefore, by means of the closed range theorem, the kernel of the adjoint of $\nabla$, i.e., ${\mathcal{H}^{\frac{3}{2}}(\Sigma,\, \nabla^\star_k p^k_a = 0,\, vol_\Sigma)}^k_a$, is a closed split subspace of ${\mathcal{H}^{\frac{3}{2}}(\Sigma,\, vol_\Sigma)}^k_a$ whose orthogonal complement coincide with the image of $\nabla$, say ${\nabla \mathcal{H}^{\frac{3}{2}}(\Sigma,\, vol_\Sigma)}^a_k$.
That is, the following splitting into closed (and, thus, Hilbert) subspaces exists:
\be 
{\mathcal{H}^{\frac{3}{2}}(\Sigma,\, vol_\Sigma)}^k_a \,=\, {\mathcal{H}^{\frac{3}{2}}(\Sigma,\, \nabla^\star_k p^k_a = 0,\, vol_\Sigma)}^k_a \oplus \nabla {\mathcal{H}^{\frac{5}{2}}(\Sigma,\, vol_\Sigma)}^a_k \,,
\ee
and the $p^k_a$'s of $\mathcal{M}_\infty$ lie exactly in the first component of such splitting.
Therefore, we can conclude that $\mathcal{M}_\infty$ is the following Hilbert space:
\be
\mathcal{M}_\infty \,=\, \prod_{k,a} \mathcal{H}^{\frac{5}{2}}_a(\Sigma,\, vol_\Sigma)^a_k \times \prod_{k,a} \mathcal{H}^{\frac{3}{2}}(\Sigma,\, \nabla^\star_k p^k_a = 0,\, vol_\Sigma) \,,
\ee
whose tangent space at each point reads:
\be
\mathbf{T}_{(a,p)}\mathcal{M}_\infty \,=\, \prod_{k,a} \mathcal{H}^{\frac{5}{2}}_a(\Sigma,\, vol_\Sigma)^a_k \times \prod_{k,a} \mathcal{H}^{\frac{3}{2}}(\Sigma,\, \nabla^\star_k p^k_a = 0,\, vol_\Sigma) \,.
\ee
Such tangent space also coincide with the space of solutions of the linearization of the constraint $\nabla^\star_k p^k_a \,=\,0$, i.e., with the space of functions ${\mathbb{X}_a}^a_k$ and ${\mathbb{X}_p}^k_a$ (representing the components of the tangent vector) satisfying:
\be
\nabla^\star_k {\mathbb{X}_p}^k_a \,=\, [p^k,\, {\mathbb{X}_a}_k]_a \,,
\ee
as the following proposition proves.
\begin{proposition}
The space of solutions of:
\be \label{Eq:linearization constraint}
\nabla^\star_k {\mathbb{X}_p}^k_a \,=\, [p^k,\, {\mathbb{X}_a}_k]_a \,,
\ee
is an affine space modelled over the vector space $\mathcal{H}^{\frac{3}{2}}(\Sigma,\, \nabla^\star_k p^k_a = 0,\, vol_\Sigma)$.
\begin{proof}
As we proved above, $\mathcal{H}^{\frac{3}{2}}(\Sigma, vol_\Sigma)$ splits as $\mathcal{H}^{\frac{3}{2}}(\Sigma, vol_\Sigma) \,=\, \mathcal{H}^{\frac{3}{2}}(\Sigma, \nabla^\star_k {\mathbb{X}_p}_a^k = 0, vol_\Sigma) \oplus \nabla \mathcal{H}^{\frac{5}{2}}(\Sigma, vol_\Sigma)$.
Let us denote by ${\tilde{\mathbb{X}}_p}_a^k$ and $\nabla_k \mathbb{X}^a_\phi$ the components of the $p_k$-component of $\mathbb{X}$ in such a splitting.
Then, equation \eqref{Eq:linearization constraint} reads:
\be \label{Eq: particular solution}
\Delta \mathbb{X}_\phi^a \,=\, [p^k,\, {\mathbb{X}_a}_k]_a \,,
\ee
where $\Delta$ is the covariant Laplacian.
The last equation has a unique solution for any fixed ${\mathbb{X}_a}_k^a$ given by the action of the Green function of $\Delta$ on the right hand side.
This means that solutions of \eqref{Eq:linearization constraint} are parametrized by all the ${\tilde{\mathbb{X}}_p}^k_a$ (belonging to $\mathcal{H}^{\frac{3}{2}}(\Sigma, \nabla^\star_k {\mathbb{X}_p}_a^k \,=\, 0, vol_\Sigma)$) and by a particular solution of \eqref{Eq: particular solution}, i. e., it is an affine space modelled over the vector space $\mathcal{H}^{\frac{3}{2}}(\Sigma, \nabla^\star_k {\mathbb{X}_p}_a^k \,=\, 0, vol_\Sigma)$.
\end{proof}
\end{proposition}

Now, on the final manifold of the PCA, the equation:
\be
i_{\mathbb{\Gamma}_\infty} \os_\infty \,=\, \dd \mathcal{H}_\infty \,,
\ee
where $\os_\infty \,=\, \mathfrak{i}^\star_\infty \os \,:=\, \mathfrak{i}^\star_1 \os$, $\mathcal{H}_\infty \,=\, \mathfrak{i}^\star_\infty \mathcal{H} \,:=\, \mathfrak{i}_1^\star \mathcal{H}$ is well posed for a $\mathbb{\Gamma}_\infty \in \mathfrak{X}(\mathcal{M}_\infty)$.


\subsection{Poisson bracket on the solution space via coisotropic embedding}
\label{Subsec:Poisson bracket on the solution space via coisotropic embedding}

Following what we said in Sect. \ref{Subsec:The canonical pre-symplectic structure and the canonical formalism near the boundary} the idea is to use the structure $\os_\infty$ to define a Poisson bracket on the solution space of the theory.
However, as we saw above, $\os_\infty$ is pre-symplectic.
Indeed, $\mathbb{X}_\psi^{\textsc{gauge}} \in {\mathbf{T}_{\chi_\Sigma} \mathfrak{i}_1 (\mathcal{M}_\infty)}^\perp$ and, furthermore, it is actually tangent to $\mathfrak{i}_\infty(\mathcal{M}_\infty)$, that is, it is $\mathfrak{i}_\infty$-related with a tangent vector to $\mathcal{M}_\infty$ which has, again, $a_k^a$ component equal to $\nabla_k \psi^a$ and $p^k_a$ component equal to $[p^k, \psi]_a$ and which we will still denote by $\mathbb{X}_\psi^{\textsc{gauge}}$.
Therefore, in order to define a Poisson bracket on $\mathcal{M}_\infty$, we may use the coisotropic embedding theorem as explained in Sect. \ref{Subsec:Poisson brackets on pre-symplectic manifolds}.
In particular, the idea is to use a well known connection on the bundle $\mathcal{M}_\infty \to \mathcal{M}_\infty / \mathrm{ker}\os_\infty$, i.e., the so-called \textit{Coulomb connection} introduced in \cite{Narasimhan1979-SU(2), Singer1978-Gribov_ambiguity} to perform the construction outlined in Sect. \ref{Sec:The coisotropic embedding theorem and Poisson brackets}.
Such a connection is represented by the following $1-1$ tensor field over $\mathcal{M}_\infty$:
\be \label{Eq:coulomb connection}
\mathcal{P} \,=\, P^a_k \otimes \frac{\delta}{\delta a^a_k} + [p^k, \mathscr{G}]_a \otimes \frac{\delta}{\delta p^k_a} \in \mathscr{T}^1_1(\mathcal{M}_\infty) \,,
\ee
where in the tensor product an integration over $\Sigma$ is implied and where:
\be
P^a_k \,=\, \nabla_k \int_\Sigma G_{\Delta} \eta^{lj} \nabla_l \delta a^a_j vol_\Sigma \,, \qquad \mathscr{G}^a \,=\, \int_\Sigma G_{\Delta} \eta^{lj} \nabla_l \delta a^a_j \, vol_\Sigma \,,
\ee
with $G_\Delta$ being the Green function of the covariant Laplacian opertator $\Delta \,=\, \eta^{jk} \nabla_j \nabla_k$ and $\left\{\,\delta a^a_j,\, \delta p^j_a \, \right\}$ being a basis of differential one forms dual to the basis of vector fields $\left\{\, \frac{\delta}{\delta a^a_k},\, \frac{\delta}{\delta p^k_a} \,\right\}_{a=1,...,dim\mathfrak{g}, k=1,2,3}$.
Note that, \eqref{Eq:coulomb connection} is actually a connection on the bundle $\mathcal{M}_\infty \to \mathcal{M}_\infty / \mathrm{ker}\os_\infty$ because it is the identity on vertical tangent vectors, i.e.:
\be
\mathcal{P}(\mathbb{X}_\psi^{\textsc{gauge}}) \,=\, \mathbb{X}_\psi^{\textsc{gauge}} \,,
\ee
and, as it is proven in \cite{Narasimhan1979-SU(2), Singer1978-Gribov_ambiguity}, it is equivariant with respect to the vertical automorphisms of the bundle.
We will denote by $\mathcal{R} \,:=\, \mathbb{1} - \mathcal{P}$ the projector over horizontal vector fields.
The latters are indeed defined to be the image of $\mathcal{R}$, say $\mathrm{Im}\mathcal{R}$.

Now, in order to apply the procedure outlined in Sect. \ref{Sec:The coisotropic embedding theorem and Poisson brackets} we should identify the dual of the vector space spanned at each point of $\mathcal{M}_\infty$ by $\mathbb{X}_\psi^{\textsc{gauge}}$.
By looking at the expression of $\mathbb{X}^{\textsc{gauge}}_\psi$, the subspace of $\mathbf{T}_{(a,p)} \mathcal{M}_\infty$ spanned by $\mathbb{X}^{\textsc{gauge}}_\psi$ at $(a, p)$, is parametrized by the $\psi^a \in {\mathcal{H}^{\frac{7}{2}}(\Sigma,\, vol_\Sigma)}^a$ and is actually the subspace of $\mathbf{T}_{(a,p)} \mathcal{M}_\infty$ given by the image of the operator $\nabla \oplus [p^k,\, \cdot \,]$ acting on  ${\mathcal{H}^{\frac{7}{2}}(\Sigma,\, vol_\Sigma)}^a$:
\be
\mathscr{V} \,:=\, \prod_{k, a} \nabla {\mathcal{H}^{\frac{7}{2}}(\Sigma,\, vol_\Sigma)}^a_k \times [p^k,\, \mathcal{H}^{\frac{7}{2}}(\Sigma,\, vol_\Sigma)]_a \,,
\ee
where $\nabla$ is the covariant derivative associated with the fixed connection $a$ of the point $(a,p) \in \mathcal{M}_\infty$.
That $\mathscr{V}$ is a Hilbert space itself, is a consequence of the fact that  $\nabla \oplus [p^k,\, \cdot \,]$ is a closed operator acting on ${\mathcal{H}^{\frac{7}{2}}(\Sigma,\, \nabla,\, vol_\Sigma)}^a$.
Indeed, the following two propositions hold.
\begin{proposition}[\textsc{Closedness of $\nabla$ in $\mathcal{H}^{\frac{5}{2}}$}]
$\nabla$ is a closed operator from ${\mathcal{H}^{\frac{7}{2}}(\Sigma,\, vol_\Sigma)}^a$ into ${\mathcal{H}^{\frac{5}{2}}(\Sigma, \, vol_\Sigma)}^a_k$.
\begin{proof}
The proof is analogous to the of Prop. \ref{Prop: closedness nabla in H3/2}.
\end{proof}
\end{proposition}
\begin{proposition}[\textsc{Closedness of $[p^k,\, \cdot \,]$}]
$[p^k,\, \cdot \,]$ is a closed operator from ${\mathcal{H}^{\frac{7}{2}}(\Sigma,\, vol_\Sigma)}^a$ to $\mathcal{H}^{\frac{7}{2}}(\Sigma,\, \nabla^\star_k p^k_a \,=\, 0,\, vol_\Sigma)$.
\begin{proof}
This is a consequence of the discussion between Eq. \eqref{Eq:commutator scalar product} and \eqref{Eq:norm commutator} 
\end{proof}
\end{proposition}

Being $\mathscr{V}$ a Hilbert space, it has a well defined dual space (isomorphic with $\mathscr{V}$ itself). 
Let us denote it by $\mathscr{V}^\star$ let us denote by ${\mathbb{X}_\mu}^k_a$ its elements.
Then, the construction of Sect. \ref{Subsec:Poisson brackets on pre-symplectic manifolds} can be applied.
In particular, since $\dd P^a_k$ is different from zero for the connection chosen here, we are in the case described in Sect. \ref{Subsubsec:The non-closed case}.
The symplectic manifold one constructs here is a tubular neinghborhood of the zero section of the bundle $\mathbf{K}^\star$ over $\mathcal{M}_\infty$ with typical fibre $\mathscr{V}^\star$.
Denote by $\tilde{\mathcal{M}}$ such a manifold and by $(a_k^a,\, p^k_a,\, \mu_a^k)$ a system of local coordinates defined over an open set $U_{\tilde{\mathcal{M}}} \subset \tilde{\mathcal{M}}$.
Then the symplectic structure on $\tilde{\mathcal{M}}$ reads:
\be
\tilde{\Omega} \,=\, \tau^\star \os_\infty\bigr|_{\mathrm{Im}\mathcal{R}} + \dd {\mu_a}^k_a \wedge P^a_k + \dd {\mu_p}^k_a \wedge [p^k,\, \mathscr{G}]_a + \mu^k_a \dd P^a_k \,,
\ee
where $\tau$ is the projection from $\tilde{\mathcal{M}}$ to $\mathcal{M}_\infty$.
Then, given two functions on $\mathcal{M}_\infty$, say $f$ and $g$, the structure $\tilde{\Omega}$ above, allows to write the Poisson bracket between $\tilde{f}\,:=\,\tau^\star f$ and $\tilde{g}\,:=\, \tau^\star g$, which reads:
\be
\{\,\tilde{f},\, \tilde{g}\,\} \,=\, \tilde{\Omega}(\mathbb{X}_{\tilde{f}},\, \mathbb{X}_{\tilde{g}}) \,,
\ee
where $\mathbb{X}_{\tilde{f}}$ and $\mathbb{X}_{\tilde{g}}$ are the Hamiltonian vector fields associated with $\tilde{f}$ and $\tilde{g}$ via $\tilde{\Omega}$.

As a specific example, consider the following functions on $\mathcal{M}_\infty$:
\be
f \,=\, p^{k_1}_{a_1}(\underline{x}_1) \, \qquad g \,=\, p^{k_2}_{a_2}(\underline{x}_2) \,.
\ee
Their pull-back to $\tilde{\mathcal{M}}$ reads:
\be
\tilde{f} \,=\, p^{k_1}_{a_1}(\underline{x}_1) \, \qquad \tilde{g} \,=\, p^{k_2}_{a_2}(\underline{x}_2) \,.
\ee
Then, a direct computation shows that:
\be
\begin{split}
\left\{ \tilde{g}, \tilde{f} \right\} \,=\, \mathbb{X}_{\tilde{f}} (\tilde{g}) \,=\, -2 \int_\Sigma \biggl[\, &{\mu_a}_a^k(\underline{y}) \nabla_k^{\underline{y}} G_\Delta(\underline{y}, \underline{x}) \delta^{k_1 k_2} \epsilon^a_{\ a_1 a_2} \delta(\underline{x}, \underline{x}_1) \delta(\underline{x}, \underline{x}_2) + \\
+ &{\mu_p}^a_k(\underline{y}) G_\Delta(\underline{y}, \underline{x}) p^k_b(\underline{y}) \delta^{k_1 k_2} \epsilon_a^{\ bc} \epsilon_{c a_1 a_2} \delta(\underline{x}, \underline{x}_1) \delta(\underline{x}, \underline{x}_2) \,\biggr] \dd^3 x \dd^3 y \,.
\end{split}
\ee

\section*{Conclusions}

In the present manuscript we concluded the program, started in the companion paper \cite{Ciaglia-DC-Ibort-Marmo-Schiav-Zamp2021-Cov_brackets_toappear}, of constructing a Poisson bracket on the space of the solutions of the equations of motion (always referred to as \textit{solution space}) of a large class of classical field theories, namely, first order Hamiltonian ones.

To resume, in this series of papers we showed how, within the multisymplectic formulation of first order Hamiltonian field theories, the solution space is canonically equipped with a pre-symplectic $2$-form.

In particular we saw that for some theories (analysed in \cite{Ciaglia-DC-Ibort-Marmo-Schiav-Zamp2021-Cov_brackets_toappear}) it is actually (strongly) symplectic and, thus, it gives rise automatically to a Poisson bracket given by the bivector field being the inverse of the symplectic structure.

Moreover, we saw that for those theories exhibiting a gauge symmetry, the canonical $2$-form has a non-trivial kernel.
We started the analysis of gauge theories in \cite{Ciaglia-DC-Ibort-Marmo-Schiav-Zamp2021-Cov_brackets_toappear} and concluded it in the present paper.
In particular we saw that, being the $2$-form pre-symplectic, via the coisotropic embedding theorem a symplectic, and, thus, a Poisson, structure on a suitable enlargement of the solution space can be always defined.
Moreover, we saw that for those theories for which a global gauge fixing in the space of fields can be performed, such Poisson structure projects to a Poisson structure on the solution space, while for those theories exhibiting Gribov's ambiguities the use of additional degrees of freedom, interpreted as ghost fields, is necessary in order to define a Poisson structure.
What is more, we classified these two cases via a geometrical structure, that is a connection on the characteristic bundle associated with the solution space equipped with the pre-symplectic structure.
In particular the two cases are selected by the connection being flat or not.

The definition of a Poisson bracket on the solution space of Einstein's equations within the Palatini's formulation of General Relativity via the approach adopted in the present paper will be addressed in a subsequent paper.

\bibliographystyle{alpha}
\bibliography{Biblio}

\end{document}